\documentclass{article}

\usepackage{microtype}
\usepackage{graphicx}
\usepackage{subfigure}
\usepackage{booktabs} 

\usepackage{hyperref}
\usepackage[utf8]{inputenc}
\usepackage{amsmath, amssymb, amsthm, bbm, mathrsfs}
\usepackage{pifont}
\newcommand{\cmark}{\ding{51}}%
\newcommand{\xmark}{\ding{55}}%
\usepackage{verbatim}
\usepackage{multirow}
\usepackage{booktabs, array, threeparttable}
\usepackage{rotating}
\usepackage{graphicx}
\usepackage{subfigure}
\usepackage{float}
\usepackage{caption}
\usepackage{amsfonts}
\usepackage{xr}
\usepackage{enumitem}
\usepackage{pgf, tikz}
\usetikzlibrary{shapes,decorations,arrows,calc,arrows.meta,fit,positioning}
\tikzset{
    -Latex,auto,node distance =1 cm and 1 cm,semithick,
    state/.style ={circle, draw, minimum width = 1.5 cm},
    point/.style = {circle, draw, inner sep=0.04cm,fill,node contents={}},
    bidirected/.style={Latex-Latex, dashed},
    el/.style = {inner sep=2pt, align=left, sloped}
}

\usepackage{etoolbox}
\usepackage{tabularx, booktabs, graphicx}

\makeatletter
\newcommand{\ostar}{\mathbin{\mathpalette\make@circled\star}}
\newcommand{\make@circled}[2]{%
  \ooalign{$\m@th#1\smallbigcirc{#1}$\cr\hidewidth$\m@th#1#2$\hidewidth\cr}%
}
\newcommand{\smallbigcirc}[1]{%
  \vcenter{\hbox{\scalebox{0.77778}{$\m@th#1\bigcirc$}}}%
}
\makeatother

\newcommand{\lt}{\left}
\newcommand{\rt}{\right}

\newcommand{\commenting}[1]{}

\renewcommand{\hat}{\widehat}

\newcommand{\Var}[1]{{\operatorname{Var}\left\{#1\right\}}}
\newcommand{\Cov}[2]{{\operatorname{Cov}\left\{#1,#2\right\}}}

\newcommand{\E}[1]{{\bbE\left\{#1\right\}}}
\newcommand{\Prob}[1]{{\bbP\left\{#1\right\}}}

\newcommand{\ind}[1]{\boldsymbol{1}\left\{#1\right\}}

\newcommand{\indep}{\perp \!\!\! \perp}

\ifdef{\see}{\renewcommand{\see}[1]{\text{ (#1)}}}{\newcommand{\see}[1]{\text{ (#1)}}}

\def\boxit#1{\vbox{\hrule\hbox{\vrule\kern6pt\vbox{\kern6pt#1\kern6pt}\kern6pt\vrule}\hrule}}

\newcolumntype{P}[1]{>{\centering\arraybackslash}p{#1}}
\newcolumntype{M}[1]{>{\centering\arraybackslash}m{#1}}
\newcolumntype{L}[1]{>{\raggedright\arraybackslash}m{#1}}

\newcommand{\cA}{{\mathcal{A}}}

\newcommand{\cN}{{\mathcal{N}}}

\newcommand{\cF}{{\mathcal{F}}}

\newcommand{\bbP}{{\mathbb{P}}}

\newcommand{\bbR}{{\mathbb{R}}}

\newcommand{\bbE}{{\mathbb{E}}}
\newcommand{\bbN}{{\mathbb{N}}}

\newcommand{\bzero}{{\boldsymbol{0}}}

\newcommand{\bI}{{\boldsymbol{I}}}

\newcommand{\hmu}{{\hat{\mu}}}

\newcommand{\hp}{{\hat{p}}}

\newcommand{\hQ}{{\hat{Q}}}

\renewcommand{\Prob}[2][]{\bbP_{#1}\lt\{#2\rt\}}
\renewcommand{\E}[2][]{\bbE_{#1}\lt\{#2\rt\}}
\renewcommand{\Var}[2][]{\operatorname{Var}_{#1}\lt\{#2\rt\}}
\renewcommand{\Cov}[3][]{\operatorname{Cov}_{#1}\lt\{#2,#3\rt\}}

\newtheorem{condition}{Condition}[section]

\usepackage[accepted]{icml2023}

\usepackage{amsmath}
\usepackage{amssymb}
\usepackage{mathtools}
\usepackage{amsthm}

\theoremstyle{plain}
\newtheorem{theorem}{Theorem}[section]

\newtheorem{lemma}[theorem]{Lemma}
\newtheorem{corollary}[theorem]{Corollary}
\theoremstyle{definition}

\theoremstyle{remark}
\newtheorem{remark}[theorem]{Remark}

\usepackage[textsize=tiny]{todonotes}

\icmltitlerunning{Statistical Inference on Multi-armed Bandits with Delayed Feedback}

\begin{document}

\twocolumn[
\icmltitle{Statistical Inference on Multi-armed Bandits with Delayed Feedback}




\begin{icmlauthorlist}
\icmlauthor{Lei Shi}{A}
\icmlauthor{Jingshen Wang}{A}
\icmlauthor{Tianhao Wu}{B}
\end{icmlauthorlist}

\icmlaffiliation{A}{Division of Biostatistics, University of California, Berkeley, USA}
\icmlaffiliation{B}{Department of Electrical Engineering and Computer Sciences, University of California, Berkeley, USA}

\icmlcorrespondingauthor{Lei Shi}{leishi@berkeley.edu}

\icmlkeywords{Machine Learning, ICML, Delayed Bandit, Statistical Inference, Hajek Estimation}

\vskip 0.3in
]



\printAffiliationsAndNotice{}  

\begin{abstract}
Multi armed bandit (MAB) algorithms have been increasingly used to complement or integrate with A/B tests and randomized clinical trials in e-commerce, healthcare, and policymaking. Recent developments incorporate possible delayed feedback. While existing MAB literature often focuses on maximizing the expected cumulative reward outcomes (or, equivalently, regret minimization), few efforts have been devoted to establish valid statistical inference approaches to quantify the uncertainty of learned policies. We attempt to fill this gap by providing a unified statistical inference framework for policy evaluation where a target policy is allowed to differ from the data collecting policy, and our framework allows delay to be associated with the treatment arms. We present an adaptively weighted estimator that on one hand incorporates the arm-dependent delaying mechanism to achieve consistency, and on the other hand mitigates the variance inflation across stages due to vanishing sampling probability. In particular, our estimator does not critically depend on the ability to estimate the unknown delay mechanism. Under appropriate conditions, we prove that our estimator converges to a normal distribution as the number of time points goes to infinity, which provides guarantees for large-sample statistical inference. We illustrate the finite-sample performance of our approach through Monte Carlo experiments.
\end{abstract}

\section{Introduction}

\subsection{Motivation and contribution}

In recent years, multi armed bandit (MAB) algorithms have been frequently used to complement A/B tests and clinical trials in practice, potentially because MAB algorithms not only aim to identify the best policies but also improve the overall outcomes for participants enrolled in the experiments. Whenever participant outcomes (or feedback) are not immediately observed, an increasing number of recent advancements further expand the practicality of classical MAB algorithms by incorporating such random \textit{delays}. While carrying out MAB algorithms in real world scenarios can be time consuming and labor intensive, there is an increased desire to be able to use those adaptively collected data from MAB algorithms to assist future decision making by answering the following cause and effect questions: Does one content recommendation plan lead to more revenue than others in e-commerce for consumers who are not enrolled in the experiment? Does one medical treatment plan cause better clinical outcomes than other plans in clinical trials? 

To answer the above questions, following the Neyman-Rubin causal model \citep{neyman1923application, rubin1974estimating}, we shall first formalize the causal parameters of interest and establish nonparametric identification results with arm-dependent \textit{delayed} feedback (Section \ref{Sec:problem-setup}), meaning that the unobserved causal effect can be written as a function of the observed data. We then build a unified statistical inference framework allowing us to construct consistent point estimates and valid confidence intervals on the causal parameters in the presence of \textit{delayed} outcomes when the number of time point goes to infinity (Section \ref{Sec:Inference-framework}). This inference framework thus enables us to answers those raised questions with rigorous statistical guarantees. In what follows, we briefly summarize our contributions:

From a methodological standpoint, on the one hand, we propose a consistent causal effect estimator that converges to a normal distribution under a wide range of unknown delay mechanisms without estimating the arm-delay joint density (Theorem \ref{thm:consistency}). As a result, the proposed estimator avoids estimating the delay mechanisms and can be more reliable compared to estimators using estimated arm-delay joint densities with noisy nonparametric approaches. On the other hand, our framework alleviates a well-known tension between the MAB design objective (regret minimization or reward maximization) and statistical inference objectives. This is achieved by simultaneously adaptive reweighting under-sampled arms similar to \citet{luedtke2016statistical, hadad2021confidence} and self-normalizing the propensity score weights with inflated variance. The tension exists because MAB algorithms are often designed to maximize expected cumulative reward outcomes and tend to assign all participants to a beneficiary arm, which leaves limited evidence to compare between the expected outcomes of a beneficiary arm and a seemingly inferior arm.
The two above features of our approach together allows us to construct valid confidence intervals of the desired causal parameters in the presence of delayed feedback and vanishing propensity scores. 

From a theoretical point of view, under appropriate conditions, we provide guarantees of the proposed statistical inference framework by proving that when the number of time points goes to infinity: (1) the proposed estimator converges the true causal effect in probability (Theorem \ref{thm:consistency}), (2) our estimator converges to a normal distribution (Theorem \ref{thm:AN}), and (3)  the variance of our proposed estimator can be consistently estimated (Theorem \ref{thm:var-est}). The adopted sufficient conditions reveal the trade-offs among the tails of the delay distribution, the outcome distribution and the vanishing propensity score distribution. Furthermore, to solidate our high level conditions, we use $\epsilon$-greedy as an illustrative sample to demonstrate the feasibility of our framework in Section \ref{sec:exp-greedy}.  In particular, our theoretical result reveals that our estimator is reliable in the sense that, establishing asymptotic normality of our estimator requires neither the propensity score of any arm to converge to a positive constant, nor the estimated outcome model to converge to the true expected outcome. This is a new contribution of our approach compared to the existing literature that adopts adaptive weighting strategy.

From a practical point of view, combining our propose statistical inference framework with rigorous large sample guarantees, we hope that our approach can be readily used to assist future decision making. Take e-commerce for example, practitioners may adopt our approach to provide an accurate revenue estimate of a content recommendation plan with confidence intervals and conduct hypothesis testing to decide if two plans lead to significant revenue difference. 

\subsection{Related literature}

Our approach is built upon the data collection mechanism in delayed multi-armed bandit algorithms. \citet{li2019bandit} studied bandit online learning with unknown delays. \citet{vernade2020linear} proposed learning algorithms for linear Bandits with stochastic delayed feedback.  
\citet{gael2020stochastic} investigated the setting of stochastic bandits with arm-dependent delays. \citet{lancewicki2021stochastic} studied stochastic bandits with unrestricted delay distributions. These works provided estimators that work well for non-vanishing propensity scores but can fail  when the sampling probability is small. \citet{zhou2019learning} dug into generalized linear contextual bandits in the presence of stochastic delays.
\citet{gyorgy2021adapting} adapted bandit learning algorithms to accommodate data in adversarial MABs. In MAB algorithms, the data are typically collected following certain learning algorithms, such as $\epsilon$-greedy, upper confidence bound (UCB) methods and Thompson sampling, to name a few \citep{sutton1998introduction}. These learning programs provide a sequence of running policies that are history-dependent and evolves adaptively with time. Other than multi-armed bandit problems, delays are commonly encountered in the practical RL literature \citep{schuitema2010control,liu2014impact,mahmood2018setting,derman2021acting}, but is less studied in theory. Recently, \citet{howson2021delayed} studied regret minimization in episodic Markov decision processes with stochastic delays. \citet{lancewicki2022learning} studied MDPs with adversarial delays under full-information feedback. \citet{jin2022near} further proposed an online-learning style algorithm for MDPs with adversarial delays under bandit feedback. 

Conducting statistical inference on datasets collected from MAB algorithms has attracted attention in the past few years. For example, \citet{zhang2020inference} studied statistical inference for batched bandits. \citet{hadad2021confidence} studied online policy evaluation in adaptive experiments. \citet{zhan2021off, bibaut2021post, chen2021statistical, zhou2022offline} studied off-policy evaluation and inference in contextual bandits. \citet{zhang2021statistical} proposed inference strategies based on M-estimation for adaptively collected data. \citet{shi2020statistical} studied inference for off-policy evaluation in reinforcement learning settings. \citet{shi2022off} studied inference for confounded markov decision processes. \citet{dimakopoulou2021online} discussed online bandits with adaptive inference.   \citet{ramprasad2022online} proposed online bootstrap inference for policy evaluation in reinforcement learning settings.  \citet{han2022online} studied online matrix contextual bandit setting with subgradient descent methods. Nevertheless, inferring delayed bandits is still a large missing piece across the statistical literature.

\section{MAB algorithms with delayed feedback: Problem setup}\label{Sec:problem-setup}

We start with introducing the structure of adaptively collected data from MAB algorithms with delayed feedback. For each time point $t \in [T] \triangleq \{1, \ldots, T\} $, one action $A_{t}$ is taken according to an underlying policy (see definition in Condition \ref{condition:policy-historical dependence}). Due to the possible delay of the feedback, we may not be able to observe the outcome $Y_{t}$ immediately. Instead, we need to wait for a certain period of time $D_{t} \in \bbN_\infty$ to acquire the outcome $Y_{t}$. In other words, if $D_{t} = 0$, we have access to $(A_{t}, Y_{t})$; otherwise, only $A_{t}$ is available at time point $t$, while $Y_{t}$ will be visible at time $t + D_{t}$. In our context, we stress that even with unknown delays, the observed rewards are identifiable in the sense that the agent knows which observed reward corresponds to which played action, even if observed with delays. To clearly characterize the causal parameters of interest, in accordance with the Neyman-Rubin causal model \citep{neyman1923application, rubin1974estimating}, we denote by $Y_{t}(a)$ the unobserved potential outcome that would be observed if action $a$ was taken at time point $t$. Given $K$ actions $\cA\triangleq \{a_1, \ldots,a_K\}$, we denote the vector of potential outcomes indexed at time point $t$ as $\{Y_t(a)\}_{a\in\cA}\in \bbR^{K}$.

For the observed outcome, we work under a frequently adopted stable unit treatment value assumption (SUTVA) in the causal inference literature \citep{imbens2015causal}:
\begin{condition}[SUTVA]\label{cond:SUTVA} The outcome at time point $t$ is determined by  action $A_t$ and not impacted by potential outcomes at other time points, suggesting that the observed outcome at $t$ to be formulated as $Y_{t} = Y_{t}(A_{t})$. 
\end{condition}

Define historical data $H_{t}$ as follows:
\begin{align*}
H_0 = \varnothing; ~H_{t} = \{(A_{t'}, (Y_{t'})_{a\in\cA}, D_{t'})\}_{t'\le t} \text{ for } t\ge 1.
\end{align*}

For the policy $\pi_t$ (with subscript $t$) taken at time point $t$, we assume:
\begin{condition}[Historical dependence of policy]\label{condition:policy-historical dependence}
$\pi_t(\cdot)$ is a vector in $\Delta(\cA)$ and only depends on the historical data. Here $\Delta(\cA)$ is a probability simplex for an action space $\cA = \{a_1,\cdots, a_K\}$
\end{condition}
Following the mainstream literature on causal inference \citep{imbens2015causal}, we refer to $\pi_t(a)$ as the propensity score of arm $a$.

\begin{figure}[!ht]
\centering
\resizebox{\columnwidth}{!}{
\begin{tikzpicture}
    \node[state] (dots1) at (-4, 0) {$H_{t-1}$};
    \node[state, dashed] (y) at (1.5,2) {$Y_{t}$};
    \node[state, dashed] (d) at (-1.5,2) {$D_{t}$};
    \node[state] (a) at (0,0) {$A_{t}$};
    
    \path (dots1) edge (a);
    \path (a) edge (d);
    \path (a) edge (y);

    \node[draw=blue, dotted, fit=(a)(y)(d), inner sep=0.2cm] (box1) {};

    \node[state] (dots2) at (4, 0) {$\cdots$}; 
    \path (d) edge (dots2);
    \path (y) edge (dots2);
    \path (a) edge (dots2);
    \path (dots1) edge[bend right=50] (dots2);
\end{tikzpicture}
}
\caption{Data generating mechanism for MAB with  delay-independent feedbacks}
\label{fig:dgp-dependent}
\end{figure}
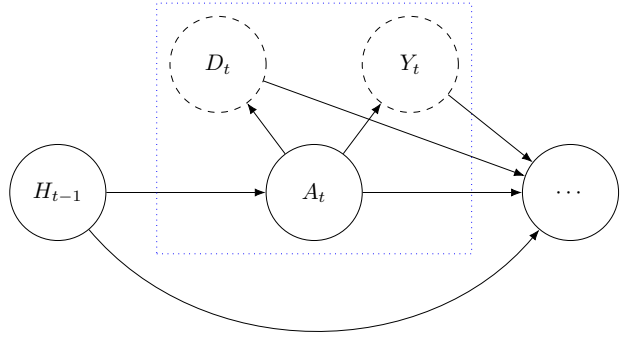

Throughout this manuscript, we assume that collected data are generated following the directed cyclic graph (DAG) in Figure \ref{fig:dgp-dependent}. 
To be more rigorous, we also formalize the causal relationships in Figure \ref{fig:dgp-dependent} with mathematical languages in Conditions \ref{cond:action-delay} and \ref{cond:PO}. Concretely, for the actions and delays, we assume:
\begin{condition}[Action-delay mechanism]\label{cond:action-delay} The action-delay pairs are generated as follows:
\begin{enumerate}[label = (\roman*)]
   \item Each $A_{t}$ is generated according to distribution $\pi_t$.
   \item Each $D_{t}$ is generated according to conditional distribution given the action $A_t$:
   \begin{align*}
     \Prob{D_t = d|A_t=a, H_{t-1}} \triangleq \Prob[a]{D = d},\  d\in \bbN_{\infty}.
   \end{align*}
\end{enumerate}
\end{condition}

For the actions, potential outcomes and delays, we assume:
\begin{condition}[Distribution of potential outcomes]\label{cond:PO}
For $t\in[T]$, $\{Y_{t}(a)\}_{a\in\cA}$ are assumed to be generated as i.i.d. copies of a random vector $\{Y(a)\}_{a\in\cA}$ which follows an unknown distribution $\bbP$. Besides, for any $t\in[T]$, $\{Y_{t}(a_1),\dots,Y_{t}(a_K)\}$ is independent of the action and delay at step $t$: 
$$
\{Y_{t}(a_1),\dots, Y_t(a_K)\} \indep  (A_{t},D_{t}).
$$
\end{condition}

We first note that Condition \ref{cond:action-delay} does not restrict the delays to be independent of actions, which broadens the applicability of our proposed framework in practice. For example, in clinical trials, our framework allows the delay mechanisms to differ across different treatment plans. Furthermore, we allow the delay to take an infinity value $+\infty$ and hence allow a part of the participants to be censored from the system. This is referred to as ``partially observed MAB'' in the literature \citep{chapelle2014modeling, krishnamurthy2009partially}. Lastly, the above conditions also imply that observed outcome at time point $t$ is independent of delay conditional on the historical data and the action, in the sense that:
\begin{align*}
Y_{t} \indep D_{t} \mid (H_{t-1}, A_t).
\end{align*}




\section{Inference target and proposed method}\label{Sec:Inference-framework}

In this section, we formulate our inference target following the problem setup introduced in the previous section and propose a  unified statistical inference framework that simultaneously accounts for unknown delay mechanism (i.e., unknown $\Prob[a]{D = d}$) and vanishing propensity scores (i.e., $\pi_t(a)\rightarrow 0$ as $t\rightarrow \infty$). 

To unify presentations, we aim to make inference on the following causal parameter: 
\begin{align}\label{eqn:parameter}
    Q(w^\star) &= \E{\sum_{a\in\cA}w^\star(a)Y(a)}= \sum_{a\in\cA}w^\star(a)Q(a), 
\end{align}
where $w^\star(\cdot) \in \bbR$ is a pre-specified function of $a\in\mathcal{A}$, and $Q(a) = \mathbb{E}[Y(a)] $ measures the expected potential outcome of arm $a$. 

The above general parameter of interest $Q(w^\star)$ is a unified presentation of several causal parameters of interest in practice. For example, when we set $w^\star(a)=1$ and  $w^\star(a')=0$ for $a'\neq a$, $Q(w^\star) = Q(a)$ which evaluates the causal effect of arm $a$. When we compare the causal effect sizes between two arms $a_1$ and $a_2$, we can set $w^\star(a_1)=1$,  $w^\star(a_2)=-1$ and $w^\star(a)=0$ for all other $a$ which does not equal to $a_1$ and $a_2$. As another example, when we want to conduct policy evaluation where a target policy $\pi$ is allowed to differ from the data collection policy $\{\pi_t\}_{t=1}^T$, then we can define $w^\star(a) = \pi(a)$. Here, a policy $\pi$ (without subscript $t$) is an element of $\Delta(\cA)$ which is a probability simplex for an action space $\cA = \{a_1,\cdots, a_d\}$. 


The causal parameter defined in \ref{eqn:parameter} can not be ``nonparametrically identified" in its current form, because it involves unobserved potential outcomes. Here, by ``nonparametric identification" we mean that the causal parameter involving unobserved potential outcomes can be written as a function of observed data. Without the delayed outcome, classical identification approaches such as inverse propensity score weighting (IPW, see \citet{rosenbaum1983central}) enable us to identify $Q(w^\star)$ with 
\begin{align}\label{eqn:identification}
    Q(w^\star) = \E{\sum_{a'\in\cA}\frac{w^\star(a)Y_t\ind{A_t = a}}{\pi_t(a)}}.
\end{align}
The intuition for the above identification result is to reweight observations using the sampling probability $\pi_t(a)$ of each arm $a$. 

In the presence of delayed feedback, the above classical IPW based identification approach in \ref{eqn:identification} is no longer valid, simply because $Y_t$ might not be available at time point $t$ due to delay. To address this issue, we first provide a new identification result appropriately taking the delay mechanism into account: 
\begin{align*}
    Q(w^\star)= & \E{\sum_{a\in\cA}\frac{w^\star(a)Y_t\ind{A_t = a, D_t \le T-t}}{\pi_t(a)\Prob[a]{D_t\le T-t}}}.
\end{align*}
Because the expectation on the right hand side of the above equation only involves observed data, as long as the arm-delay joint density $\Prob[a]{D_t\le T-t}$ is well estimated, it seems that an estimator of $Q(w^\star)$ can be naturally constructed by replacing the unknown expectation with its sample analogue. 

Nevertheless, two practical challenges potentially occur when directly using the above plug-in approach. On the one hand, accurately estimation of unknown conditional delayed distribution either requires imposing potentially misspecified parametric models on $\Prob[a]{D_t\le T-t}$ or depends on nonparametric approaches with potentially noisy behaviors in practice. Moreover, estimating $\Prob[a]{D_t\le T-t}$  can be particularly challenging when the delays present a heavy-tailed pattern. As a remedy, we aim to propose an estimator of $ Q(w^\star)$ avoids estimating of delaying distribution either parametrically or nonparametrically, and thus free us from the burden of estimating arm-delay distribution. 

On the other hand, because MAB algorithms are designed for regret minimization and inferior arms are less likely to be sampled when $t$ is large, this suggests that for those inferior arms the propensity scores $\pi_t(a)$ can be a very small number for large $t$. In the presence of vanishing propensity scores, IPW based estimators tend to have large variance and no longer converge to a normal distribution as the number of time points goes to infinity \citep{hadad2021confidence, ma2020robust}. This renders statistical inference based on standard normal approximations invalid.

To address the above challenges, we propose a ``Delay-adjusted augmented inverse propensity weighting" (DAIPW) estimator that, not only accounts for the arm-dependent delay mechanisms, but also restores the asymptotic normal distribution by self-normalizing and adaptively weighting the arms with vanishing propensity scores:
\begin{align}\label{eqn:hQ}
    \hQ_{\texttt{DAIPW}}(w^\star) = \sum_{a\in\cA} w^\star(a)\hQ_{\texttt{DAIPW}}(a),
\end{align}
where
\begin{align}\label{eqn:hQa}
    \hat{Q}_{\texttt{DAIPW}}(a) &= \frac{\sum_{t\in[T]}h_t(a)\lt\{ (Y_{t} - \hmu_t(a))\gamma_t(a)\rt\}}{\sum_{t=1}^T h_t(a) \gamma_t(a)}\notag\\
    &+ \frac{\sum_{t\in[T]}h_t(a)\hmu_t(a)}{\sum_{t\in[T]}h_t(a)}
\end{align}
and $\gamma_t(a)$ is the inverse propensity score weighted indicator of whether the outcome is delayed at time $t$ for arm $a$,
\begin{align*}
    \gamma_t(a) = \frac{\ind{A_{t} = a, D_{t}\le T-t}}{\pi_t(a)}.
\end{align*}
Here, $\hat{\mu}_t(a)$ is an estimator for the outcome model $\mu(a) = \mathbb{E}[Y_t | A_t=a ]$ at time $t$, and $h_t(a)$ is a sequence of adaptive weights that, intuitively, down-weight the role of under-sampled arms in constructing our estimator. Both $\hat{\mu}_t(a)$ and $h_t(a)$ are constructed only using the historical data $H_{t-1}$ collected before time $t$. As $h_t(a)$ plays an essential role in constructing $ \hQ_{\texttt{DAIPW}}(w^\star)$, we demonstrate in Section 4.2 how to adaptively choose $h_t(a)$ in commonly adopted MAB algorithms, where we use $\epsilon$-greedy as an illustrative example.

Before proposing a variance estimator for uncertainty quantification, we provide three additional insights hoping to demonstrate the potential merits of our estimator $ \hQ_{\texttt{DAIPW}}(w^\star)$. 

First, the proposed estimator takes the presence of delays into account by the introduction of the inverse propensity score weighted observation indicator $\gamma_t(a)$. A critical point is that \textbf{we neither need to adjust for the unknown conditional delay distribution $\Prob[a]{D_t\le T-t}$ nor need to worry about situation where the delay mechanism displays a heavy-tailed pattern. } 
Under appropriate conditions listed in Section \ref{sec:asp-theory}, we can still justify the consistency of $\hQ_{\texttt{DAIPW}}(w^\star)$ even for heavy-tailed delays.

Second, the proposed estimator can be viewed as a delay-adjusted generalization of the augmented IPW estimator (AIPW) and the classical H\'{a}jek estimator \citep{hajek1971comment}, therefore it inherits some instinct strength from these two estimators. On the one hand, due to augmenting the IPW estimator with the estimated outcome model $\hat{\mu}_t(a)$, our estimator extracts additional information stored in the data and may further improve our finite sample performance. While keeping the bias almost the same, applying the augmented estimator improves the finite sample behavior of the variance estimators. On the other hand, \textbf{the adaptation of H\'{a}jek estimation demonstrates two-fold benefits: it is both helpful for variance stabilization due to small propensity scores and necessary for adjusting for delayed, even possibly never observed, outcomes.} In fact, in the extreme case where a part of the outcomes are never observed, the non-H\'{a}jek version of our estimator estimator is seriously biased, which is demonstrated by our additional theoretical derivation and simulation results provided in Section B.1 of the Appendix. 

Third, the proposed estimator \textbf{incorporates arm-wise adaptive weights $h_t(a)$ to mitigate the variance inflation} due to possibly vanishing propensity scores (i.e., $\pi_t(a)\rightarrow 0$ as $t\rightarrow \infty$). Such a consideration is a generalization of the non-delayed policy evaluation schemes provided in \citet{luedtke2016statistical, bibaut2021post, hadad2021confidence, zhan2021off}. 

Before we end the section, we introduce a variance estimator of $\hQ_{\texttt{DAIPW}}(w^\star)$, 
\begin{align*}
 \hat{V}  & = \sum_{a\in\cA}\{w^\star(a)\}^2\hat{V}(a),\\
    \hat{V}(a) & = \frac{\sum_{t\in[T]}h_t(a)^2\lt\{(Y_t - \hQ_{\texttt{DAIPW}}(a))\gamma_t(a)\rt\}^2}{\lt(\hp(a)\sum_{t\in[T]}h_t(a)\rt)^{2}},
\end{align*}
where $\hp(a)$ is an estimator of $p(a) = \Prob[a]{D<\infty}$ (i.e., the probability of having finite delays for arm $a$)
\begin{align*}
   \hp(a) = \frac{\sum_{t\in[T]} h_t(a) \gamma_t(a)}{\sum_{t\in[T]}h_t(a)}.
\end{align*}
The point estimator $\hQ_{\texttt{DAIPW}}(w^\star)$ and the variance estimator $ \hat{V}  $ together enable us to construct $(1-\alpha)$-level confidence interval for $Q(w^\star)$:
\begin{align*}
    \Big[\hQ_{\texttt{DAIPW}}(w^\star) \pm z_{\alpha/2}{\hat{V}}^{\frac{1}{2}}  \Big],
\end{align*}
where $z_{\alpha/2}$ is the upper $(1-\alpha/2)$ quantile of a standard normal distribution. In our simulation results provided in Section \ref{sec:simulation}, we set $\alpha = 0.05$. 


\section{Theoretical investigation}\label{sec:theory}

In this section, under appropriate conditions, we provide guarantees of the proposed statistical inference framework by proving that when $T\rightarrow \infty$: (1) $\hQ_{\texttt{DAIPW}}(a)$ converges to $Q(a)$ in probability (Theorem \ref{thm:consistency}), (2) $\hQ_{\texttt{DAIPW}}(a) - Q(a)$ converges to a centered normal distribution with variance $V(a)$ (Theorem \ref{thm:AN}), and (3) $ \hat{V}(a) $ converges to $V(a)$ in probability (Theorem \ref{thm:var-est}), for all arms $a\in \mathcal{A}$. As $Q(w^\star)$ is a weighted combination of $Q(a)$, the three results above together justifies the statistical validity of the proposed framework in an asymptotic sense (i.e., $T$ is large). 

Furthermore, to solidate our high level conditions in \ref{asp:small-h}-\ref{eqn:variance-converge}, we use $\epsilon$-greedy as a running sample to demonstrate the feasility of these conditions in Section \ref{sec:exp-greedy}.


\subsection{Large sample guarantees of the proposed estimator}\label{sec:asp-theory}

We begin by providing a consistency result of our estimator in \ref{eqn:hQ}. The core idea is to decompose the error $\hQ(a) - Q(a)$ into two parts. The first part is a martingale sequence with stable variance and bounded moments under property weighting and mild conditions on the potential outcomes. The second part is an asymptotically vanishing remainder term compared to the first part. Then we can conclude the proof by carefully checking the conditions for martingale limit theorems \citep{hall1980martingale}.

We list the assumptions used to quantify the scale of the adaptive weights $h_t(a)$ as well as the interplay between $h_t(a)$, the moments of $Y(a)$ and the delay distribution. 
\begin{enumerate}[label = (A\arabic*)]
    \item \label{asp:small-h} \textit{Negligible adaptive weights.} For all $a\in\cA$,
    \begin{align}\label{eqn:small-h}
      \frac{\max_{t\in[T]} h_t(a)}{\sum_{t=1}^T h_t(a)} \xrightarrow{\bbP} 0.
    \end{align}
    \item \label{asp:ht-D} \textit{Appropriate delay tails.} For all $a\in\cA$,
    \begin{align}\label{eqn:ht-D}
    \frac{\sum_{t\in[T]}h_t(a)\Prob[a]{T-t<D<\infty}}{\lt(\sum_{t=1}^T \E{\frac{h_t(a)^2}{\pi_t(a)} \Prob[a]{D\le T-t}}\rt)^{\frac{1}{2}}} = O_\bbP(1).
    \end{align}
    \item \label{asp:inf-sample} \textit{Infinite sampling.} For all $a\in\cA$,
    \begin{align}\label{eqn:inf-sample}
       \frac{\E{\sum_{t=1}^T h_t(a)^2\pi_t(a)^{-1} \Prob[a-1]{D_t\le T-t}}}{\lt(\sum_{t=1}^T h_t(a)\rt)^2} \xrightarrow{\bbP} 0.
    \end{align}
    \item \label{asp:lyapunov} \textit{Lyapunov condition.} For all $a\in\cA$,
    \begin{align}\label{eqn:lyapunov}
      \frac{\sum_{t=1}^T h_t(a)^{2+\delta}  \pi_t(a)^{-(1+\delta)}\Prob[a]{D\le T-t}}{\lt({\sum_{t=1}^T \E{\frac{h_t(a)^2}{\pi_t(a)} \Prob[a]{D\le T-t}} }\rt)^{\frac{2+\delta}{2}}} \xrightarrow{\bbP} 0,
    \end{align}
    \item \label{asp:variance-converge} \textit{Variance convergence condition.} For all $a\in\cA$, these exists some $p>1$, 
    \begin{align}\label{eqn:variance-converge}
       \frac{{\sum_{t=1}^T {h_t(a)^2 \pi_t(a)^{-1}\Prob[a]{D\le T-t}} }}{{\sum_{t=1}^T \E{h_t(a)^2 \pi_t(a)^{-1}\Prob[a]{D\le T-t}} }} \xrightarrow{L_p} 1.
    \end{align}
\end{enumerate}

We add some comments on these conditions. Condition \ref{asp:small-h} requires each single adaptive weight is negligible compared to the sum of all the weights. Intuitively speaking, this ensures that the weights are relatively balanced in magnitude and there is no dominating components or outliers. Condition \ref{asp:ht-D} requires the tail of the delay $\Prob[a]{T-t < D <\infty}$ to vanish at some appropriate rate. Later we will show that Condition \ref{asp:ht-D} allows heavy tailed delay which may even have bounded expectation. Condition \ref{asp:inf-sample} is standard in the adaptive-weighting based policy evaluation frameworks \citep[e.g.][]{hadad2021confidence}. We generalize it to incorporate delay distributions. Condition \ref{asp:lyapunov} and \ref{asp:variance-converge} are needed for martingale limit theories. In general, Condition \ref{asp:small-h} to \ref{asp:variance-converge} can be used as criteria or guidance to construct valid adaptive weights. In Section \ref{sec:exp-greedy} we elaborate more on these conditions using one concrete data collection MAB algorithms. 

Under Conditions \ref{asp:small-h} to \ref{asp:variance-converge}, we next shows the consistency of our estimator:

\begin{theorem}[Consistency]\label{thm:consistency}
Under Conditions \ref{asp:small-h} - \ref{asp:lyapunov}, we further assume that $Y(a)$ has $(2+\delta)$-th moment for some small positive $\delta$ and $\hmu_t(a)$ is bounded and converges in probability to some constant. We have 
\begin{align*}
    \hat{p}(a) - p(a) \xrightarrow{\bbP} 0, \quad \hQ_{\texttt{\em DAIPW}}(a) - \mu(a) \xrightarrow{\bbP} 0.
\end{align*}
\end{theorem}

Theorem \ref{thm:consistency} consists of two results: consistency of the estimated non-censoring probability $\hp(a)$ and consistency of the DAIPW estimator $\hQ_{\texttt{DAIPW}}(a)$ for each arm. Then former part works as an intermediate result which provides evidence and insights of the validity of DAIPW even in the presence of never observed outcomes. It is also a crucial component for justifying the validity of the variance estimation in Theorem \ref{thm:var-est}.

We next present the theoretical result demonstrating that our estimator converges to a normal distribution when $T$ goes to infinity: 
\begin{theorem}[Asymptotic normality]\label{thm:AN}
Under Conditions \ref{asp:small-h} - \ref{asp:variance-converge}, we further assume that $Y(a)$ has $(2+\delta)$-th moment for some small positive $\delta$ and $\hmu_t(a)$ is bounded and converges in probability to some constant. We have 
\begin{align*}
    \lt[\frac{\hQ_{\texttt{\em DAIPW}}(a_k) - \mu(a_k)}{V(a_k)^{1/2}}\rt]_{k\in[K]}^{\top} \to \cN(\bzero, \bI_K),
\end{align*}
where 
\begin{align*}
    V(a) = \frac{\sum_{t\in[T]}\E{h_t(a)^2\pi_t(a)^{-1}\sigma_a^2\Prob[a]{D\le T-t}}}{\lt(p(a)\sum_{t\in[T]}h_t(a)\rt)^{2}}.
\end{align*}
\end{theorem}

Theorem \ref{thm:AN} implies the proposed estimators $[\hQ(a)]_{a\in\cA}$ follow a multivariate normal distribution as $T$ goes to infinity. Besides, their asymptotic between-arm correlation is zero. As $Q(w^\star)$ is a linear combination of $Q(a)$, this suggests that the proposed estimator $ \Big[\hQ_{\texttt{DAIPW}}(w^\star)$ also converges to a normal distribution. 

A direct implication of Theorem \ref{thm:AN} is that, \textbf{the asymptotic normality of our estimator requires neither the propensity scores $\pi_t(a)$ on any arm to converge to a positive constant, nor the $\hat{\mu}(a)$ to converge to the true $\mathbb{E}[Y_t|A_t = a]$}. This unique property of our estimator is a result of incorporating H\'{a}jek-type strategy in MAB problems, which is a new contribution of our approach compared to the existing literature that adopts adaptive weighting strategy.


Lastly, we justify the validity of the variance estimation and complete the story of statistical inference.

\begin{theorem}[Variance estimation]\label{thm:var-est}
Under the same conditions listed in Theorem \ref{thm:AN}, we have 
\begin{align*}
    \frac{\hat{V}(a)}{V(a)} \xrightarrow{\bbP} 1.
\end{align*}
\end{theorem}

\subsection{Verification of our framework in $\epsilon$-greedy algorithms}\label{sec:exp-greedy}

In this section, we discuss the choice of adaptive weights in the $\epsilon$-greedy algorithms and show that the proposed weights satisfy Condition \ref{asp:small-h}-\ref{asp:variance-converge}.

In $\epsilon$-greedy algorithms, the algorithm picks the arm $a_{t,\max}$ that achieves the highest sample average. Then in the following stage the algorithm pulls $a_{t,\max}$ with probability greater than $1-\epsilon$ (exploitation) and the rest $(d-1)$ arms with probability $\epsilon/(d-1)$ (exploration). In practice, we can set a diminishing $\epsilon_t$ to reduce the portion of exploration. In particular, we study policy evaluation problem of the power decaying $\epsilon_t$; i.e., $\epsilon_t = t^{-\alpha}$ for some $\alpha\ge 0$. Mathematically, let $\overline{Y}_{a, t}$ denote the  averaged observed outcome collected on arm $a$ before time $t$. Then
\begin{gather}\label{eqn:epsilon-greedy}
    \pi_t(a) = \left\{
    \begin{array}{cc}
       1 - \epsilon_t,  &  \overline{Y}_{a, t-1} \ge
       \overline{Y}_{a', t-1};\\
       \frac{\epsilon_t}{d-1},  & 
       \overline{Y}_{a, t-1} < \overline{Y}_{a', t-1}.
    \end{array}
    \right.
\end{gather}

A simple adaptive weighting strategy is available in this algorithm by setting
\begin{align}\label{eqn:choose-ht}
    h_t(a) = \sqrt{\pi_t(a)}.
\end{align}
In the non-delayed setting, this was studied by \citet{hadad2021confidence} and termed as ``constant allocation strategy''.  The following corollary justifies this choice in our current setup with delays:
\begin{corollary}\label{cor:epsilon-greedy}
    Assume  $\pi_t(a) \ge  Ct^{-\alpha}$  for some $\alpha \in [0,1)$. Also assume the delay distribution satisfies:
\begin{gather*}
    \Prob[a]{D = 0} >0, \\
    \Prob[a]{t \le  D < \infty} = O(t^{-\beta}) \text{ for all } t\ge 1 \text{ and some } \beta\ge \frac{1}{2}.
\end{gather*}
Then \ref{asp:small-h} - \ref{asp:variance-converge} are satisfied. 
\end{corollary}

\begin{remark}
    Define the margin of the bandit as the gap between the largest and second largest expected outcome. The requirement of $ \beta \ge 1/2 $ can be relaxed to $\alpha + \beta \ge \frac{1}{2}$ if the margin is nonzero. See the proof in the Appendix for more details. 
\end{remark}

We can check that for the $\epsilon$-greedy algorithm, the conditions in Corollary \ref{cor:epsilon-greedy} can be satisfied. More generally, Corollary \ref{cor:epsilon-greedy} can be applied for a wider class of online learning algorithms, such as Thompson sampling or UCB-based algorithms (see Section B.2 of the Appendix). The sole requirement is that the algorithm preserves certain level of probability for randomization through the trajectory ($\pi_t(a) \ge Ct^{-\alpha}$ where $\alpha$ is not too small). Such conditions have been discussed similarly in many other sequential policy evaluation frameworks, such as batched bandit learning \citep{zhang2020inference},  etc. 

\section{Simulation study}\label{sec:simulation}

In this section, we verify our theoretical results in two simulation designs by comparing the performance of the following estimators in Table \ref{tab:estimators}. More concretely, \texttt{DAIPW} is the proposed estimator. \texttt{Mean} is the simple sample mean estimator $\hat{Q}(a) = N(a)^{-1}\sum_{t=1}^TY_t\boldsymbol{1}(A_t = a, D_t\le T-t)$. \texttt{NH0} is the ordinary (non-H\'{a}jek) IPW estimator $\hat{Q}(a) = T^{-1}\sum_{t=1}^TY_t\boldsymbol{1}(A_t = a, D_t\le T-t)/\pi_t(a)$. \texttt{NH} is the usual AIPW estimator by \citet{hadad2021confidence}, i.e.,  \texttt{NH0} with outcome adjustment.
\begin{table}[!ht]
\centering
\caption{Estimators Considered in Numerical Studies}
\label{tab:estimators}
\resizebox{\columnwidth}{!}{%
\begin{tabular}{ccccc}
\hline
\textbf{Estimator}           & \texttt{DAIPW}   & \texttt{Mean}   & \texttt{NH}         & \texttt{NH0}   \\ \hline
\textbf{Delay-adjusted}      & \cmark & \cmark & \xmark     & \xmark \\
\textbf{Adaptively-weighted} & \cmark & \xmark & \cmark     & \cmark \\
\textbf{H\'{a}jek-type}      & \cmark & \cmark & \xmark     & \xmark \\
\textbf{Outcome-adjusted}    & \cmark & \xmark & \cmark & \xmark \\ \hline
\end{tabular}%
}
\end{table}
In the first simulation design, we study the performance of varies estimators with/without margins. In the second simulation design, we compare different estimators under multiple delay mechanisms.

\subsection{Simulation results with different margins in $\epsilon$-greedy algorithms}

\begin{figure}[!ht]
    \centering
    \includegraphics[width = \columnwidth]{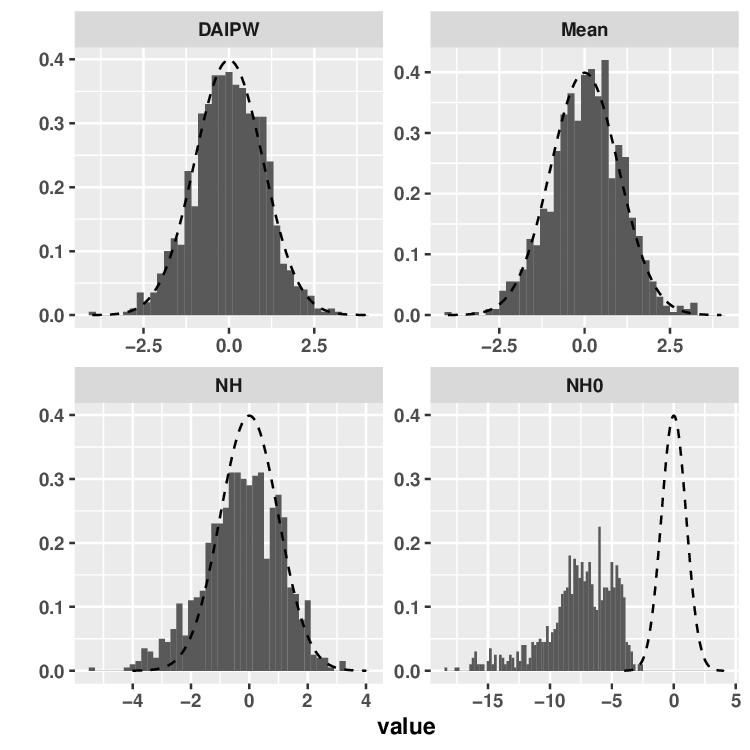}
    \caption{$\epsilon$-greedy over zero margin bandits $\mu(1) - \mu(2) = 0$. }
    \label{fig:zero-margin}
\end{figure}

\begin{figure}[!ht]
    \centering
    \includegraphics[width = \columnwidth]{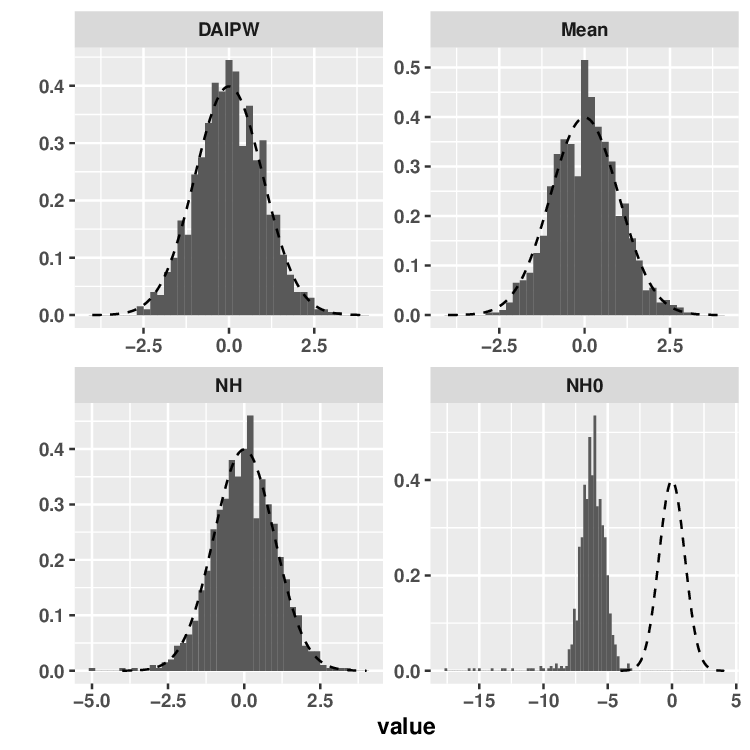}
    \caption{$\epsilon$-greedy over nonzero margin bandits with  $\mu(1) - \mu(2) = 0.1$. }
    \label{fig:nonzero-margin}
\end{figure}

We run $\epsilon$-greedy on binary bandits and evaluate the impact of margins for the performance of varies estimators in Table \ref{tab:estimators}. The potential outcomes are generated from normal distribution with variance 1. For Arm $a=1$, there exists a positive censoring probability: $\Prob[1]{D = \infty} = 0.5$. Arm 2 does not have never observed outcome, i.e., the censoring probability  $\Prob[2]{D = \infty} = 0$. We compare two settings with different sizes of margins: (i) zero margin: $\mu(1) - \mu(2) = 0$; (ii) non-zero margin: $\mu(1) - \mu(2) = 0.1$. The results are summarized in Figure \ref{fig:zero-margin} and Figure \ref{fig:nonzero-margin}.

From Figure \ref{fig:zero-margin} and Figure \ref{fig:nonzero-margin} we can see that, the proposed \texttt{DAIPW} estimator provides a better approximation the normal distribution in both zero margin and nonzero margin settings. The non-adaptively weighted estimator with delay adjustment (\texttt{Mean}) slightly skewed to the left for the non-zero margin case. The skewing effect will be more prominent as the margin increases since the propensity score $\pi_t(2)$ converges to 0 faster. The non-H\'{a}jek estimator (\texttt{NH}) works poorly when the margin is zero, because the outcome model estimator converges slower in that scenario. \texttt{NH0}, which neither accounts for delays nor adjusts for the outcome model, is severely biased regardless of the size of the margin. 

\subsection{Simulation results under different delay mechanisms}

In this section, we run $\epsilon$-greedy on a binary bandit with $\mu(1) = 1.0$ and $\mu(2) = 0.5$.  We compare four different delay mechanisms:
\begin{itemize}
    \item No finite delay. No other source of delay is included except for the censoring on Arm 1. 
    \item Negative binomial delay. The delay distribution on both arms follow Negative Binomial distributions, which gives a subexponential-tailed delay.
    \item Pareto delay. The delay distribution on both arms follow (rounded) Pareto distributions, which gives a polynomial-type heavy-tailed delay.
\end{itemize}

\begin{figure}[!ht]
    \centering
    \includegraphics[width = \columnwidth]{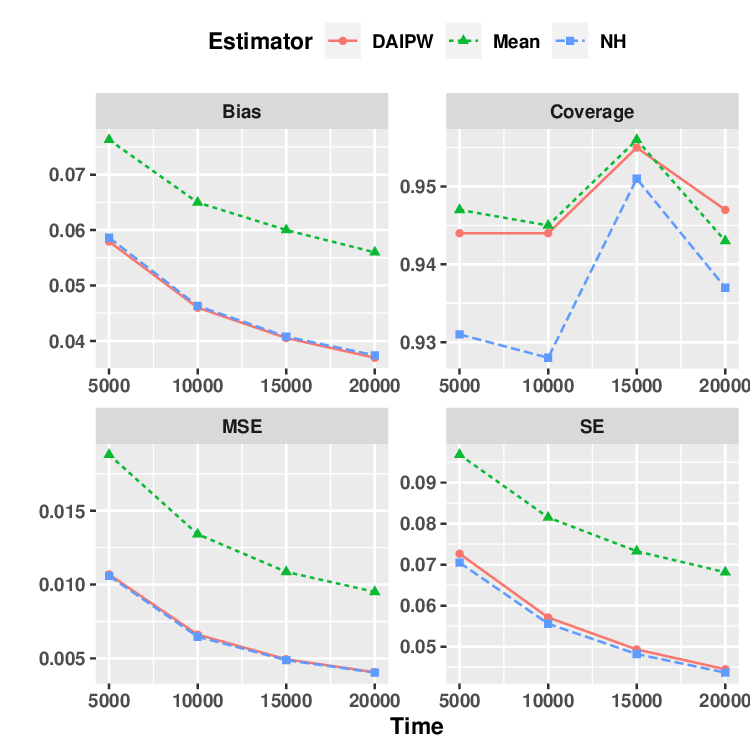}
    \caption{Evaluation of $\epsilon$-greedy with no finite delays}
    \label{fig:no-delay}
\end{figure}

Due to space limit, we present the first case with no finite delay, and leave the simulation results for other settings to the Supplementary Material. The results are summarized in Figure \ref{fig:no-delay}.
From Figure \ref{fig:no-delay}, we can see that the unweighted estimator (\texttt{Mean}) has a higher absolute bias, and the estimated standard deviation based on \texttt{Mean} is also larger, which leads to a longer confidence interval. Using the proposed adaptive weighted H\'{a}jek estimator (\texttt{DIPW}) or adjusting for the outcome model (\texttt{NH}) can mitigate the problem. However, the non-H\'{a}jek estimator (\texttt{NH}) tends to underestimate the variance because it does not adjust for the censoring probability, leading to under covered confidence intervals. In sum, our approach is the only method that provides accurate point estimate and valid confidence interval (meaning that the coverage probability attains the nominal 95\% level).


\section{Discussion}

In this manuscript, we provide a unified statistical inference framework when data are adaptively collected from MABs with delayed feedback. Under appropriate conditions, we prove that our estimator converges to a normal distribution as the number of time points goes to infinity, which provides guarantees for large-sample statistical inference. 

We add some discussions on the potential generalizations and challenges of our framework. First,
a realistic generalization can potentially include delay-dependent outcome or outcome-dependent delay mechanism. Nonparametric identification and statistical inference can be more challenging in these settings. Second, practitioners may also care about statistical quantification of optimal policies, which is unknown and are typically learned through online algorithms.  Therefore, it is of interest to establish methodology that combines policy evaluation and policy learning. Third, it is interesting to unify two multi-armed bandit setting across the literature: batched bandit (small $T$ and large sample $n_t$ per batch) and the current long-horizon bandit (large horizon $T$ and small sample $n_t$ at each $t$). Finite sample probability bounds such as Berry-Esseen type results will be helpful for such unification \citep{hall1980martingale, shi2022berry}. We leave these possible extensions as future endeavor.



\section*{Software}

The source code is available from the GitHub repository: \url{https://github.com/LeiShi-rocks/DelayBandits}. 

\section*{Acknowledgements}

We are very thankful for the questions and comments of the three anonymous reviewers.


\bibliography{ref}
\bibliographystyle{icml2023}

\newpage
\appendix
\onecolumn
\section{Proof of the main results}
\subsection{Some probability results}

The following lemma is attributed to Theorem 3.2 and Corollary 3.1 of \citet{hall1980martingale}, which provides a central limit theorem for martingale difference arrays:

\begin{lemma}[CLT for martingale difference array]
Let $\{S_{ni}, \cF_{ni}, 1\le i\le k_n ,n\ge 1\}$ be a zero-mean, square-integrable martingale array with differences $X_{ni}$ and let $\eta^2$ be an a.s. finite r.v. Suppose that the following conditions hold:
\begin{enumerate}[label = (\roman*)]
    \item Conditional Lindeberg condition:
    \begin{align*}
        \text{for all } \epsilon > 0, \sum_{i\in[k_n]} \E{X_{ni}^2\ind{|X_{ni}| > \epsilon} \mid \cF_{n, i-1}}\xrightarrow{\bbP} 0,
    \end{align*}
    \item Convergence of conditional variance:
    \begin{align*}
        W^2_{nk_n} = \sum_{i\in[k_n]} \E{X^2_{ni}\mid \cF_{n,i-1}} \xrightarrow{\bbP} \eta^2.
    \end{align*}
    \item Nested $\sigma$-fields condition:
    \begin{align*}
        \cF_{n,i} \subset \cF_{n+1, i} \text{ for } 1\le i \le k_n, n\ge 1. 
    \end{align*}
\end{enumerate}
\end{lemma}

Two important quantities in the study of a (zero-mean) martingale 
$$
\{S_{ni} = \sum_{j\in[i]}X_{nj}, \cF_{ni}, 1\le i \le k_n\}
$$ 
is the \textit{conditional variance} and the \textit{squared variation}, which are both estimation of the variance $\E{S_n^2}$ and are defined as follows:
\begin{align*}
    W_{ni}^2 &= \sum_{j\in[i]} \E{X_{nj}^2\mid \cF_{n,j-1}},\\
    U_{ni}^2 &= \sum_{j\in[i]} X_{nj}^2,~ \text{ for } 1\le i \le k_n. 
\end{align*}

The following lemma is taken from Theorem 2.23 of \citet{hall1980martingale}, which gives conditions under which $V_{ni}^2$ and $U_{ni}^2$ are asymptotically equivalent:
\begin{lemma}[Asymptotic equivalence of $W_{ni}^2$ and $U_{ni}^2$]\label{lem:asp-equivalent}
Suppose the following conditions hold:
\begin{enumerate}[label = (\roman*)]
    \item The conditional variances $W_{nk_n}^2$ are tight:
    \begin{align*}
        \sup_{n} \Prob{W_{nk_n}^2 > \lambda} \to 0 \text{ as } \lambda \to \infty. 
    \end{align*}
    \item The conditional Lindeberg condition holds: 
    \begin{align*}
        \forall \epsilon > 0, \sum_{i\in[k_n]} \E{X_{ni}^2\ind{|X_{ni}| > \epsilon} \mid \cF_{n, i-1}}\xrightarrow{\bbP} 0,
    \end{align*}
\end{enumerate}
Then 
\begin{align*}
    \max_i |U_{ni}^2 - W_{ni}^2| \xrightarrow{\bbP} 0.
\end{align*}
\end{lemma}

A sufficient condition for the conditional Lindeberg condition is the so called \textit{conditional Lyapunov condition}: for some $\delta>0$, 
\begin{align*}
\sum_{i\in[k_n]}\E{|X_{n,i}|^{2+\delta}\mid\cF_{n,i-1}} \xrightarrow{\bbP} 0.
\end{align*}

We cite a Lemma from \citet{hadad2021confidence} (Lemma 10), which is inherently a probabilistic version of the well-known Topelitz lemma \citep{hall1980martingale}:
\begin{lemma}[Relaxed version of Lemma 10 of \citet{hadad2021confidence}]\label{lem:Topelitz}
Let $a_{T,t}$ be a triangular sequence of nonnegative
weight vectors satisfying 
\begin{gather*}
    \max_{1\le t\le T} a_{T,t} \xrightarrow{\bbP} 0,\quad \sum_{t=1}^T a_{T,t} = O_\bbP(1).
\end{gather*}
Let $x_t$ be a sequence of bounded random variables (with bound $B>0$) satisfying $x_t \xrightarrow{\bbP} 0$. Then 
\begin{align*}
    \sum_{t=1}^T a_{T,t} x_t \xrightarrow{\bbP} 0.
\end{align*}

\end{lemma}
\begin{remark}
Lemma 10 of \citet{hadad2021confidence} assumes $\text{\em plim}_{T\to\infty} \sum_{t=1}^T a_{T,t} \le C $ and $x_t \xrightarrow{a.s.} 0$, which is not necessary by slightly modifying the proof.
\end{remark}

\begin{proof}
For any $\epsilon>0$ and $\delta>0$, there exists $C_1$ and $T_1$, such that for any $T\ge T_1$,
\begin{align*}
    \Prob{\sum_{t=1}^T a_{T,t} < C_1} \ge 1 - \frac{\delta}{3}.
\end{align*}
There exists $T_2$, such that for any $T\ge T_2$
\begin{align*}
    \Prob{|x_t| < \frac{\epsilon}{2C_1}} > 1- \frac{\delta}{3}.
\end{align*}
There exists $T_3$, such that for any $T\ge T_3$,
\begin{align*}
    \Prob{\max_{1\le t\le T} a_{T,t} \le \frac{\epsilon}{2BT_2}} > 1 - \frac{\delta}{3}.
\end{align*}
Therefore, for any $T\ge \max\{T_1, T_2, T_3\}$, over the intersection of the above events,
\begin{align*}
    \sum_{t=1}^T a_{T,t} |x_t| &= \sum_{t=1}^{T_2-1} a_{T,t} |x_t| + \sum_{t=T_2}^T a_{T,t} |x_t| \\
    & < 
    BT_2\max_{t\in[T]}a_{T,t} + (\sum_{t=T_2}^T a_{T,t}) \frac{\epsilon}{2C_1} < \epsilon.
\end{align*}

\end{proof}

\subsection{Proof of Theorem \ref{thm:consistency}}
\begin{proof}[Proof of Theorem \ref{thm:consistency}]  
We finish the proof by combining the following three steps: (i) proving consistency of delaying probability estimation to its expectation; (ii) proving consistency of the estimator.

\noindent\textbf{Step (i).} We show that 
\begin{align}\label{eqn:hat-p}
    \hat{p}(a) \xrightarrow{\bbP} p(a).
\end{align}
\eqref{eqn:hat-p} follows from the fact that
\begin{align*}
    h_t(a) (\frac{\ind{A_{t} = a, D_{t}\le T-t}}{\pi_t(a)} - \Prob[a]{D\le T-t})
\end{align*}
is a martingale difference sequence and 
the assumption \eqref{eqn:ht-D} and \eqref{eqn:inf-sample},
\begin{align}
    \hp(a) - p(a) &= \frac{\sum_{t=1}^T h_t(a) (\frac{\ind{A_{t} = a, D_{t}\le T-t}}{\pi_t(a)} - \Prob[a]{D\le T-t})}{\sum_{t\in[T]}h_t(a)} \\
    &+ \frac{\sum_{t\in[T]}h_t(a)\Prob[a]{T-t<D<\infty}}{\sum_{t\in[T]}h_t(a)} \notag\\
    &= O_\bbP\lt(\frac{[\sum_{t=1}^T \E{h_t(a)^2 \gamma_t(a)^2}]^{1/2}}{\sum_{t\in[T]}h_t(a)}\rt) = o_\bbP(1). \label{eqn:order-hp}
\end{align}

\noindent\textbf{Step (ii).} 
We show that
\begin{align*}
    \hQ_T(a) - \mu(a) \xrightarrow{\bbP} 0.
\end{align*}
We have
\begin{align*}
    &\hQ_T(a) - \mu(a) = \\
    &\frac{\sum_{t\in[T]}h_t(a)\lt\{ (Y_{t}(a) - \mu(a) + \mu_\infty(a) - \hmu_t(a)) \gamma_t(a)\rt\}}{\sum_{t=1}^Th_t(a) \gamma_t(a)}\\
    +& \frac{\sum_{t\in[T]}h_t(a)(\hmu_t(a)-\mu_\infty(a))}{\sum_{t\in[T]}h_t(a)}.
\end{align*}
Introduce the notation
\begin{gather*}
    \gamma_t(a) = \frac{\ind{A_t = a, D_t\le T-t}}{\pi_t(a)}, \\
    \Delta_t(a) = (Y_{t}(a) - \mu(a) + \mu_\infty(a) - \hmu_t(a)) \gamma_t(a).
\end{gather*}
Then 
\begin{align}\label{eqn:two-parts-hQ}
    & \frac{\sum_{t\in[T]}h_t(a)}{p(a)^{-1}}\cdot\frac{\hQ_T(a) - \mu(a)}{[\sum_{t\in[T]}\E{h_t(a)^2(\Delta_t(a) - \E[t-1]{\Delta_t(a)})^2}]^{1/2}} \\
    &= \frac{\sum_{t\in[T]}h_t(a)\{\hp(a)^{-1}\Delta_t(a) + \hmu_t(a) - \mu_\infty(a)\}}{p(a)^{-1}[\sum_{t\in[T]}\E{h_t(a)^2(\Delta_t(a) - \E[t-1]{\Delta_t(a)})^2}]^{1/2}}\\
    &= \frac{\hp(a)^{-1}}{p(a)^{-1}}\underbrace{\frac{\sum_{t\in[T]}h_t(a)\{\Delta_t(a) - \E[t-1]{\Delta_t(a)}\}}{[\sum_{t\in[T]}\E{h_t(a)^2(\Delta_t(a) - \E[t-1]{\Delta_t(a)})^2}]^{1/2}}}_{\text{Part I}} \\
    &+ \frac{\hp(a)^{-1}}{p(a)^{-1}}\underbrace{\frac{\sum_{t\in[T]}h_t(a)\{\E[t-1]{\Delta_t(a)} 
    + (\hmu_t(a) - \mu_\infty(a))\hp(a)\}}{[\sum_{t\in[T]}\E{h_t(a)^2(\Delta_t(a) - \E[t-1]{\Delta_t(a)})^2}]^{1/2}}}_{\text{Part II}}
\end{align}
where $\E[t-1]{\Delta_t(a)} = (\mu_\infty(a) - \hmu_t(a))\Prob[a]{D\le T-t}$.

It's clear that $\text{Part I} = O_\bbP(1)$.  Now we show $\text{Part II} = o_\bbP(1)$.

For Part II, we have
\begin{align*}
    &\frac{\sum_{t\in[T]}h_t(a)\{\E[t-1]{\Delta_t(a)} 
    + (\hmu_t(a) - \mu_\infty(a))\hp(a)\}}{[\sum_{t\in[T]}\E{h_t(a)^2(\Delta_t(a) - \E[t-1]{\Delta_t(a)})^2}]^{1/2}}\\
    = &\frac{\sum_{t\in[T]}h_t(a)\{(\hmu_t(a) - \mu_\infty(a))(\hp(a) - \Prob[a]{D\le T-t})\}}{[\sum_{t\in[T]}\E{h_t(a)^2(\Delta_t(a) - \E[t-1]{\Delta_t(a)})^2}]^{1/2}}\\
    = & \frac{\sum_{t\in[T]}h_t(a)\{(\hmu_t(a) - \mu_\infty(a))(\hp(a) - p(a))\}}{[\sum_{t\in[T]}\E{h_t(a)^2(\Delta_t(a) - \E[t-1]{\Delta_t(a)})^2}]^{1/2}}\\
    + & \frac{\sum_{t\in[T]}h_t(a)\{(\hmu_t(a) - \mu_\infty(a))(p(a) - \Prob[a]{D\le T-t})\}}{[\sum_{t\in[T]}\E{h_t(a)^2(\Delta_t(a) - \E[t-1]{\Delta_t(a)})^2}]^{1/2}}.
\end{align*}

For the denominator, we first compute
\begin{align*}
    \E[t-1]{\lt|\Delta_{t}(a) - \E[t-1]{\Delta_{t}(a)}\rt|^{2}}
    = \Var[t-1]{\Delta_{t}(a)}.
\end{align*}
Now using the law of total variance, 
\begin{align*}
    \Var[t-1]{\Delta_{t}(a)} &= \E[t-1]{\Var[t-1]{\Delta_{t}(a)\mid A_{t}, D_t}} + \Var[t-1]{\E[t-1]{\Delta_t(a) \mid A_{t}, D_t}} = \text{III} + \text{IV}.
\end{align*}
For III,
\begin{align*}
    \text{III} &= \E[t-1]{\gamma_t(a)^2\Var[t-1]{Y_t(a) -\mu(a) + \mu_\infty(a) - \hmu_t(a)\mid A_{t}, D_t}}\\
    &= \E[t-1]{\gamma_t(a)^2\Var[t-1]{Y_t(a)}} = \E[t-1]{\sigma_a^2\gamma_t(a)^2}.
\end{align*}
For IV,
\begin{align*}
    \text{IV} &= \Var[t-1]{\E[t-1]{\Delta_t(a) \mid A_{t}, D_t}}\\
    &= \Var[t-1]{ (\mu_\infty(a) - \hmu_t(a))\gamma_t(a) }.
\end{align*}
Therefore,
\begin{align}
    &\E[t-1]{\lt|\Delta_{t}(a) - \E[t-1]{\Delta_{t}(a)}\rt|^{2}}\\
    = & \E[t-1]{\sigma_a^2\gamma_t(a)^2} +  \Var[t-1]{ (\mu_\infty(a) - \hmu_t(a))\gamma_t(a) }  \label{eqn:var-decomp}\\
    \ge &  \E[t-1]{\sigma_a^2\gamma_t(a)^2}.
\end{align}
Hence,
\begin{align*}
    \sum_{t\in[T]}\E{h_t(a)^2(\Delta_t(a) - \E[t-1]{\Delta_t(a)})^2} \ge \sum_{t=1}^T \E{h_t(a)^2 \sigma_a^2\gamma_t(a)^2},
\end{align*}
which gives
\begin{align*}
    \text{Part II} 
    \le & \underbrace{\frac{\sum_{t\in[T]}h_t(a)\{(\hmu_t(a) - \mu_\infty(a))(\hp(a) - p(a))\}}{[\sum_{t=1}^T \E{h_t(a)^2 \sigma_a^2\gamma_t(a)^2}]^{1/2}}}_{\text{Term I}}\\
    + & \underbrace{\frac{\sum_{t\in[T]}h_t(a)\{(\hmu_t(a) - \mu_\infty(a))(p(a) - \Prob[a]{D\le T-t})\}}{[\sum_{t=1}^T \E{h_t(a)^2 \sigma_a^2\gamma_t(a)^2}]^{1/2}}}_{\text{Term II}}.
\end{align*}

For Term I, using \eqref{eqn:order-hp}, assumption \eqref{eqn:small-h} and Lemma \ref{lem:Topelitz}, we have
\begin{align*}
    \text{Term I} = \frac{\sum_{t\in[T]}h_t(a)\{(\hmu_t(a) - \mu_\infty(a))\}}{\sum_{t\in[T]}h_t(a)}\frac{(\hp(a) - p(a))}{[\sum_{t=1}^T \E{h_t(a)^2 \sigma_a^2\gamma_t(a)^2}]^{1/2}/(\sum_{t\in[T]}h_t(a))} = o_\bbP(1).
\end{align*}

For Term II, using the Lyapunov condition \eqref{eqn:lyapunov}, we have
\begin{align}
    &\lt|\frac{\max_{t\in[T]} h_t(a)^2\E[t-1]{\gamma_t(a)^2}}{\E{\sum_{t\in[T]}h_t(a)^2 \gamma_t(a)^2}}\rt|^{\frac{2+\delta}{2}} \notag\\
    \le & \lt|\frac{\max_{t\in[T]} h_t(a)^{{2+\delta}} \E[t-1]{\gamma_t(a)^{2+\delta}}}{(\sum_{t\in[T]}\E{h_t(a)^{2}\gamma_t(a)^2})^{\frac{2+\delta}{2}}}\rt| \notag\\
    \le & \lt|\frac{\sum_{t\in[T]} h_t(a)^{{2+\delta}}\E[t-1]{\gamma_t(a)^{2+\delta}}}{(\sum_{t\in[T]}\E{h_t(a)^{2}\gamma_t(a)^2})^{\frac{2+\delta}{2}}}\rt| \xrightarrow{\bbP} 0 \see{by the Lyapunov Condition}. \label{eqn:small-wt}
\end{align}

using \eqref{eqn:ht-D}, \eqref{eqn:small-wt} and Lemma \ref{lem:Topelitz}, we can show $\text{Term II} = o_\bbP(1)$.

Therefore, Part II vanishes in probability. 

Now we combine the above results and conclude 
\begin{align*}
    \hQ_T(a) - \mu(a) = O_\bbP\lt(\frac{[\sum_{t=1}^T \E{h_t(a)^2 \gamma_t(a)^2}]^{1/2}}{\sum_{t\in[T]}h_t(a)}\rt) = o_\bbP(1).
\end{align*}

\end{proof}

\subsection{Proof of Theorem \ref{thm:AN}}

\begin{proof}[Proof of Theorem \ref{thm:AN}]
Single arm case: we hope to show
\begin{align*}
    \frac{\hQ_T(a) - \mu(a)}{V_T(a)^{1/2}} \xrightarrow{d} \cN(0,1),
\end{align*}
where 
\begin{align*}
    V_T(a) = p(a)^{-2}\sum_{t\in[T]}\E{h_t(a)^2\pi_t(a)^{-1}\sigma_a^2\Prob[a]{D\le T-t}}.
\end{align*}

We start from single-arm cases. Recall the two parts decomposition \eqref{eqn:two-parts-hQ}. We have shown that Part II converge to zero in probability under the given assumptions.

\textbf{We show that: Part I is a martingale sequence and converge to $\cN(0,1)$.}

\noindent {\textbf{Step I: show that $\xi_{t}$ is a martingale difference sequence.}}

Define 
\begin{gather*}
    \xi_{t,T}(a) = \frac{ h_t(a)\{\Delta_t(a) - \E[t-1]{\Delta_t(a)}\}}{[\sum_{t\in[T]}\E{h_t(a)^2(\Delta_t(a) - \E[t-1]{\Delta_t(a)})^2}]^{1/2}}.
\end{gather*}

It is not hard to check that the sequence of $\{\xi_{t}(a),\sigma(H_{t-1})\}_{t=1}^T$ forms a martingale difference sequence. 

\noindent {\textbf{Step II: Use martingale CLT to prove asymptotic normality.}}

The crucial step is to verify the Lyapunov condition and the variance convergence condition.

\textbf{(II-1) Lyapunov condition.}
We compute
\begin{align}\label{eqn:check-Lyapunov}
    \sum_{t=1}^T \E[t-1]{|\xi_{t,T}(a)|^{2+\delta}} &= 
    \frac{\sum_{t=1}^T h_t(a)^{2+\delta} \E[t-1]{\lt|\Delta_{t}(a) - \E[t-1]{\Delta_{t}(a)}\rt|^{2+\delta}}}{\lt[\sum_{t=1}^T  \E{h_t(a)^{2}\lt(\Delta_{t}(a) - \E[t-1]{\Delta_{t}(a)}\rt)^{2}}\rt]^{(2+\delta)/2}}.
\end{align}

(1) For the numerator, we have
\begin{align}
    \|\Delta_{t}(a) - \E[t-1]{\Delta_{t}}(a)\|_{L_{t-1}^{2+\delta}}
    \le    \|\Delta_{t}(a)\|_{L_{t-1}^{2+\delta}} + \| \E[t-1]{\Delta_{t}(a)}\|_{L_{t-1}^{2+\delta}}. \see{By Minkowski's inequality} 
\end{align}
By Jensen's inequality, we have
\begin{align*}
    \lt\| \E[t-1]{\Delta_{t}(a)}\rt\|_{L^{2+\delta}_{t-1}}  
    \le
     \E[t-1]{\lt\|{\Delta_{t}(a)}\rt\|_{L^{2+\delta}_{t-1}}}
    = 
    \lt\|{\Delta_{t}(a)}\rt\|_{L^{2+\delta}_{t-1}}.
\end{align*}
Hence
\begin{align}
     &\|\Delta_{t}(a) - \E[t-1]{\Delta_{t}(a)}\|_{L_{t-1}^{2+\delta}}\\
    & \le 2 \|\Delta_t(a)\|_{L_{t-1}^{2+\delta}} 
     = 2 \lt\|{(Y(a) - \mu(a) + \mu_\infty(a) - \hmu_{t})\gamma_{t}(a)}\rt\|_{L^{2+\delta}_{t-1}}\\
    & \le 2 \{\underbrace{\lt\|{(Y_t(a) - \mu(a))\gamma_{t}(a)}\rt\|_{L^{2+\delta}_{t-1}}}_{\text{Term I}} + \underbrace{\lt\|{(\hmu_{t}(a) - \mu_\infty(a))\gamma_{t}(a)}\rt\|_{L^{2+\delta}_{t-1}}}_{\text{Term II}}\}. \label{eqn:term-I-II}
\end{align}

For Term I in \eqref{eqn:term-I-II}, we have 
\begin{align}
    &\lt\|{(Y_t(a) - \mu(a))\gamma_{t}(a)}\rt\|^{2+\delta}_{L^{2+\delta}_{t-1}} \\
    = &\E[t-1]{\lt|{(Y_t(a) - \mu(a))\gamma_{t}(a)}\rt|^{2+\delta}} \\
    = &\E[t-1]{|\gamma_{t}(a)|^{2+\delta}\E{\lt|(Y_{t}(a) - \mu(a))\rt|^{2+\delta} \Big\vert H_{t-1}, A_{t}, D_t}} \\
    = &\E[t-1]{|\gamma_{t}(a)|^{2+\delta}\E{\lt|(Y_{t}(a) - \mu(a))\rt|^{2+\delta}}} \\
    \le& M_{2+\delta}\E[t-1]{|\gamma_{t}(a)|^{2+\delta}}.
\end{align}

For Term II in \eqref{eqn:term-I-II}, because we assumed $|\hmu(a)|$ is uniformly bounded by some $M_\mu$, Term II is also bounded (up to a constant) by $\E[t-1]{|\gamma_{t,i}(a)|^{2+\delta}}$. Therefore,
\begin{align*}
    \E[t-1]{\lt|\Delta_{t}(a) - \E[t-1]{\Delta_{t}(a)}\rt|^{2+\delta}} \le M  \E[t-1]{|\gamma_{t}(a)|^{2+\delta}}.
\end{align*}

(2) For the denominator, using the variance decomposition \eqref{eqn:var-decomp},
\begin{align*}
    &\E[t-1]{\lt|\Delta_{t}(a) - \E[t-1]{\Delta_{t}(a)}\rt|^{2}}\\
    = & \E[t-1]{\sigma_a^2\gamma_t(a)^2} +  \Var[t-1]{ (\mu_\infty(a) - \hmu_t(a))\gamma_t(a) }\\
    \ge &  \E[t-1]{\sigma_a^2\gamma_t(a)^2}.
\end{align*}

Therefore, for \eqref{eqn:check-Lyapunov}, 
\begin{align}
    \sum_{t=1}^T \E[t-1]{|\xi_{t,T}(a)|^{2+\delta}} \le \frac{M \sum_{t=1}^T h_t(a)^{2+\delta}  \E[t-1]{|\gamma_t(a)|^{2+\delta}}}{\lt({\sum_{t=1}^T \E{h_t(a)^2\sigma_a^2\gamma_t(a)^2} }\rt)^{\frac{2+\delta}{2}}} \xrightarrow{\bbP} 0.
\end{align}

\textbf{(II-2) Variance convergence.} 
Now we check variance convergence. Combining \eqref{eqn:var-decomp}, we need to show
\begin{align}\label{eqn:var-converge}
    \sum_{t=1}^T \E[t-1]{\xi_{t,T}(a)^2} = \frac{{\sum_{t=1}^T h_t(a)^2 \lt(\E[t-1]{\sigma_a^2\gamma_t(a)^2} +  \Var[t-1]{ (\mu_\infty(a) - \hmu_t(a))\gamma_t(a) } \rt) } }{\bbE\lt({\sum_{t=1}^T h_t(a)^2 \lt(\E[t-1]{\sigma_a^2\gamma_t(a)^2} +  \Var[t-1]{ (\mu_\infty(a) - \hmu_t(a))\gamma_t(a) } \rt) } \rt)} \xrightarrow{\bbP} 1.
\end{align}

Define
\begin{align*}
    Z_T &= \sum_{t=1}^T h_t(a)^2 \Bigg(\E[t-1]{\sigma_a^2\gamma_t(a)^2} \\
    &+  \Var[t-1]{ (\mu_\infty(a) - \hmu_t(a))\gamma_t(a) } \Bigg)  \\
    &  = \underbrace{\sum_{t=1}^T h_t(a)^2 \E[t-1]{\sigma_a^2\gamma_t(a)^2}}_{\text{Term I}} \\
    &+ \underbrace{\sum_{t=1}^T h_t(a)^2 \Var[t-1]{ (\hmu_t(a) - \mu_\infty(a))\gamma_t(a) }}_{\text{Term II}}.
\end{align*}

For Term I, according to the assumption \eqref{eqn:variance-converge}, we have
\begin{align*}
    \frac{\sum_{t=1}^T h_t(a)^2 \E[t-1]{\sigma_a^2\gamma_t(a)^2}}{\sum_{t=1}^T \E{h_t(a)^2 \sigma_a^2\gamma_t(a)^2}} \xrightarrow{L^p} 1.
\end{align*}

For Term II, by H\"{o}lder's inequality, 
\begin{gather*}
    |\text{Term II}|  \le \sum_{t=1}^T h_t(a)^2\E[t-1]{\gamma_t(a)^2}|\hmu_t(a) - \mu_\infty(a)|^2.
\end{gather*}
By assumption, $|\hmu_t(a) - \mu_\infty(a)|\xrightarrow{\bbP}0$. Besides, based on \eqref{eqn:small-wt}, the following weights are negligible:
\begin{align*}
    &\lt|\frac{\max_{t\in[T]} h_t(a)^2\E[t-1]{\gamma_t(a)^2}}{\E{\sum_{t\in[T]}h_t^2 \gamma_t(a)^2}}\rt|\to 0.
\end{align*}
Therefore by Lemma \ref{lem:Topelitz}, we can conclude 
\begin{align}\label{eqn:Term-II-op1}
    \frac{|\text{Term II}|}{\sum_{t=1}^T \E{h_t(a)^2 \sigma_a^2\gamma_t(a)^2}} = o_\bbP(1).
\end{align}
Combining the fact that 
\begin{align*}
    \frac{|\text{Term II}|}{\sum_{t=1}^T \E{h_t(a)^2 \sigma_a^2\gamma_t(a)^2}} \le \frac{2M_\mu^2}{\sigma_a^2},
\end{align*}
\eqref{eqn:Term-II-op1} can  be strengthened into $L^1$-convergence:
\begin{align*}
    \frac{\E{|\text{Term II}|}}{\sum_{t=1}^T \E{h_t(a)^2 \sigma_a^2\gamma_t(a)^2}} \to 0.
\end{align*}

Therefore,
\begin{align*}
    \frac{\text{Term I} + \text{Term II}}{\E{\text{Term I} + \text{Term II}}} \xrightarrow{\bbP} 0. 
\end{align*}

Combining (II-1) and (II-2), we have shown that 
\begin{align*}
    \sum_{t\in[T]} \xi_{t}(a) \xrightarrow{d} \cN(0,1).
\end{align*}

\noindent {\textbf{Step III: Justify joint asymptotic normality using Cram\'{e}r-Wold device.}}

For simplicity we consider two arms. The generalization to a finite number of arms is rather straightforward. Choose a  vector $(b_1,b_2)^\top\in\bbR^2$ with $b_1^2 + b_2^2 = 1$. It suffices to show that under the given conditions,
\begin{align*}
    b_1\xi_{t,T}(a) + b_2\xi_{t,T}(a') \xrightarrow{d} \cN(0,1).
\end{align*}
The Lyapunov condition is easy to check because we know
\begin{align*}
    &\lt|b_1\xi_{t,T}(a) + b_2\xi_{t,T}(a')\rt|^{2+\delta} \\
    \le& 2^{1+\delta}(|b_1|^{2+\delta}|\xi_{t,T}(a)|^{2+\delta} + |b_2|^{2+\delta}|\xi_{t,T}(a')|^{2+\delta}).
\end{align*}
Therefore, the Lyapunov condition follows naturally from the results for individual arms. 

The tricky part is to check variance convergence.  We briefly go through the idea of proof. 
\begin{align*}
    &\sum_{t=1}^T \E[t-1]{(b_1\xi_{t,T}(a) + b_2\xi_{t,T}(a'))^2}\\
    & = \sum_{t=1}^T b_1^2\E[t-1]{\xi_{t,T}(a)^2} + \sum_{t=1}^Tb_2^2\E[t-1]{\xi_{t,T}(a')^2} \\
     &+ \sum_{t=1}^T b_1b_2\E[t-1]{\xi_t(a)\xi_t(a')}.
\end{align*}

According to the proofs in Step 2, (II-2), 
\begin{align*}
    \E[t-1]{\xi_{t,T}(a)^2} \xrightarrow{\bbP} 1,\quad \E[t-1]{\xi_{t,T}(a')^2} \xrightarrow{\bbP} 1.
\end{align*}

Now we show
\begin{align*}
    \sum_{t=1}^T\E[t-1]{\xi_{t,T}(a)\xi_{t,T}(a')} \xrightarrow{\bbP} 0.
\end{align*}

Write down the expression explicitly, 
\begin{align*}
    &\sum_{t=1}^T\E[t-1]{\xi_{t,T}(a)\xi_{t,T}(a')}\\
    =&\frac{\sum_{t\in[T]}h_t(a)h_t(a')\Cov[t-1]{\Delta_t(a)}{\Delta_t(a')}}{[\sum_{t\in[T]}\bbE\{h_t(a)^2\Var[t-1]{\Delta_{t}(a)}\}]^{1/2}\cdot [\sum_{t\in[T]}\bbE\{h_t(a')^2\Var[t-1]{\Delta_{t}(a')}\}]^{1/2}}.
\end{align*}

For the numerator, 
\begin{align*}
    &\sum_{t\in[T]}h_t(a)h_t(a')\Cov[t-1]{\Delta_t(a)}{\Delta_t(a')}\\
    = &\sum_{t\in[T]}h_t(a)h_t(a')(\E[t-1]{\Delta_t(a)\Delta_t(a')} - \E[t-1]{\Delta_t(a)}\E[t-1]{\Delta_t(a')})\\
    = & -\sum_{t\in[T]}h_t(a)h_t(a') \E[t-1]{\Delta_t(a)}\E[t-1]{\Delta_t(a')}\\
    &\see{because $\gamma_t(a)\gamma_t(a') = 0$}.
\end{align*}
Therefore, by Cauchy-Schwarz inequality and the variance decomposition \eqref{eqn:var-decomp}, 
\begin{align*}
    &\lt|\sum_{t=1}^T\E[t-1]{\xi_{t}(a)\xi_{t}(a')}\rt|\\
    \le & \lt[\frac{\sum_{t\in[T]}h_t(a)^2(\E[t-1]{\Delta_t(a)})^2 }{\sum_{t\in[T]}\bbE\{h_t(a)^2\E[t-1]{\sigma_a^2\gamma_t(a)^2}\}}\rt]^{1/2}\\
    \cdot&\lt[\frac{\sum_{t\in[T]}h_t(a')^2(\E[t-1]{\Delta_t(a')})^2}{ \sum_{t\in[T]}\bbE\{h_t(a')^2\E[t-1]{\sigma_{a'}^2\gamma_t(a')^2}\}}\rt]^{1/2}.
\end{align*}

Now we have
\begin{align*}
    &\sum_{t\in[T]}h_t(a)^2(\E[t-1]{\Delta_t(a)})^2 \\
    \le & \sum_{t\in[T]}h_t(a)^2\E[t-1]{\gamma_t(a)^2}\frac{(\E[t-1]{\Delta_t(a)})^2}{\E[t-1]{\gamma_t(a)^2}} \\
    \le & \sum_{t\in[T]}h_t(a)^2\E[t-1]{\gamma_t(a)^2}\frac{(\E[t-1]{(\mu_\infty(a) - \hmu_t(a))\gamma_t(a)})^2}{\E[t-1]{\gamma_t(a)^2}}\\
    \le & \sum_{t\in[T]}h_t(a)^2\E[t-1]{\gamma_t(a)^2}{(\mu_\infty(a) - \hmu_t(a))^2}.
\end{align*}

Under the condition $\hmu(a) \xrightarrow{\bbP} \mu_\infty(a)$,  the variance convergence assumption \eqref{eqn:variance-converge}, negligible weights result \eqref{eqn:small-wt} and Lemma \ref{lem:Topelitz}, we can conclude:
\begin{align}
    \frac{\sum_{t\in[T]}h_t(a)^2 (\E[t-1]{\Delta_t(a)})^2 }{\sum_{t\in[T]}\bbE\{h_t(a)^2 \E[t-1]{\sigma_a^2\gamma_t(a)^2}\}} \xrightarrow{\bbP} 0.
\end{align}
\end{proof}

\subsection{Proof of Theorem \ref{thm:var-est}}
\begin{proof}[Proof of Theorem \ref{thm:var-est}]
We mainly need to show 
\begin{align}\label{eqn:var-est}
    \frac{\sum_{t\in[T]}h_t(a)^2\lt\{(Y_t - \hat{Q}_T(a))\gamma_t(a)\rt\}^2}{\sum_{t\in[T]}\E{h_t(a)^2\sigma_a^2\gamma_t(a)^2}}\xrightarrow{\bbP} 1.
\end{align}

We have the decomposition 
\begin{align*}
    &\sum_{t\in[T]}h_t(a)^2\lt\{(Y_t - \hat{Q}_T(a))\gamma_t(a)\rt\}^2 \\
    = &\underbrace{\sum_{t\in[T]}h_t(a)^2\lt\{(Y_t(a) - \mu(a))\gamma_t(a)\rt\}^2}_{\text{Term I}} \\
    + & \underbrace{\sum_{t\in[T]}h_t(a)^2\lt\{( \mu(a) - \hat{Q}_T(a))\gamma_t(a)\rt\}^2}_{\text{Term II}} \\
    + & 2\underbrace{\sum_{t\in[T]}h_t(a)^2\lt\{(Y_t(a) - \mu(a))( \mu(a) - \hat{Q}_T(a))\gamma_t(a)^2\rt\}}_{\text{Term III}}.
\end{align*}

\textbf{Term I:} Term I is a square variation of the martingale difference sequence
\begin{align*}
    \phi_{t,T}(a) = \frac{h_t(a)(Y_t(a) - \mu(a))\gamma_t(a)}{\lt[\E{\sum_{t\in[T]}{\phi_{t,T}^2}}\rt]^{1/2}}.
\end{align*}
By the variance convergence assumption \eqref{eqn:variance-converge}, the conditional square variation satisfies:
\begin{align*}
    {\sum_{t\in[T]}\E[t-1]{\phi_{t,T}^2}} \xrightarrow{\bbP} 1. 
\end{align*}
By the Lyapunov condition \eqref{eqn:lyapunov} and Lemma \ref{lem:asp-equivalent}, we have
\begin{align*}
    {\sum_{t\in[T]}{\phi_{t,T}^2}} \xrightarrow{\bbP} 1.
\end{align*}

\textbf{Term II:} for Term II, we have 
\begin{align*}
    &\frac{\sum_{t\in[T]}h_t(a)^2\lt\{( \mu(a) - \hat{Q}_T(a))\gamma_t(a)\rt\}^2}{\sum_{t\in[T]}\E{h_t(a)^2\sigma_a^2\gamma_t(a)^2}} \\
   = & ( \mu(a) - \hat{Q}_T(a))^2\frac{\sum_{t\in[T]}h_t(a)^2\lt\{\gamma_t(a)\rt\}^2}{\sum_{t\in[T]}\E{h_t(a)^2\sigma_a^2\gamma_t(a)^2}} \\
   =& o_\bbP(1) \cdot O_\bbP(1) = o_\bbP(1).
\end{align*}

\textbf{Term III:} for Term III, using Cauchy-Schwarz inequality, we have
\begin{align*}
    \text{Term III} \le (\text{Term I})^{1/2} (\text{Term II})^{1/2}.
\end{align*}
Therefore, based on the results on Term I and Term II, it's not hard to show $\text{Term III} = o_\bbP(1)$. 

Now we can combine all parts above to conclude the proof.

\end{proof}

\subsection{Proof of Corollary \ref{cor:epsilon-greedy}}
\begin{proof}[Proof of Corollary \ref{cor:epsilon-greedy}]
We check \ref{asp:small-h} to \eqref{asp:variance-converge}.

\begin{itemize}
    \item For \ref{asp:small-h}, we have
    \begin{align*}
        \frac{\max_{t\in[T]} h_t(a)}{\sum_{t=1}^T h_t(a)} \le \frac{1}{C\sum_{t\in[T]}t^{-\alpha/2}} \asymp \frac{1}{T^{1-\alpha/2}} \to 0.
    \end{align*}
    \item For \ref{asp:ht-D}, we have 
    \begin{align*}
    &\frac{\sum_{t\in[T]}h_t(a)\Prob[a]{T-t<D<\infty}}{[\sum_{t=1}^T \E{h_t^2 \pi_t(a)^{-1}\Prob[a]{D\le T-t}}]^{1/2}}\\
        = &\frac{\sum_{t\in[T]}h_t(a)\Prob[a]{T-t<D<\infty}}{[\sum_{t=1}^T {\Prob[a]{D\le T-t}}]^{1/2}}.
    \end{align*}
    For the denominator, we have
    \begin{align*}
        \sum_{t=1}^T {\Prob[a]{D\le T-t}} &= p(a) T - \sum_{t\in[T]}\Prob[a]{T-t<D<\infty} \\
        &\asymp  T - \max\{T^{1-\beta},1\} \asymp T.
    \end{align*}
    For the numerator, in general we have
    \begin{align*}
        &\sum_{t\in[T]}h_t(a)\Prob[a]{T-t<D<\infty}\\
        \le& \sum_{t\in[T]}\Prob[a]{T-t<D<\infty} \asymp T^{1-\beta}.
    \end{align*}
    Hence we need $\beta \ge 1/2$. If $\pi_t(a) \asymp t^{-\alpha}$, then the numerator can be bounded by
    \begin{align*}
        &\sum_{t\in[T]}h_t(a)\Prob[a]{T-t<D<\infty}\\
        \le &  \sum_{t\in[T]}t^{-\frac{\alpha}{2}}(T-t+1)^{-\beta} \\
        = & T^{1-\frac{\alpha}{2}-\beta} \sum_{t\in[T]}(\frac{t}{T})^{-\frac{\alpha}{2}}(\frac{T-t+1}{T})^{-\beta}\cdot\frac{1}{T}.
    \end{align*}
    When $\beta < 1/2$, 
    \begin{align*}
        &\sum_{t\in[T]}(\frac{t}{T})^{-\frac{\alpha}{2}}(\frac{T-t+1}{T})^{-\beta}\cdot\frac{1}{T}\\ \to& \int_{0}^1 x^{-\frac{\alpha}{2}}(1-x)^{-\beta}dx =  \operatorname{Beta}(1-\frac{\alpha}{2}, 1 - \beta).
    \end{align*}
    Hence we can allow $\frac{\alpha}{2} + \beta \ge \frac{1}{2} $.
    
    \item For \ref{asp:inf-sample}, we have
    \begin{align*}
        &\frac{\E{\sum_{t=1}^T h_t(a)^2\pi_t(a)^{-1} \Prob[a-1]{D_t\le T-t}}}{\lt(\sum_{t=1}^T h_t(a)\rt)^2}
        \lesssim  \frac{T}{T^{2-\alpha}} \asymp T^{-(1-\alpha)} \to 0.
    \end{align*}
    
    \item For \ref{asp:lyapunov}, the numerator satisfies
    \begin{align*}
        &\sum_{t\in[T]} h_t^{2+\delta}  \pi_t(a)^{-(1+\delta)}\Prob[a]{D\le T-t}\\
        \le& \sum_{t\in[T]} t^{\frac{\alpha\delta}{2}} \asymp T^{1 + \frac{\alpha\delta}{2}}.
    \end{align*}
    For the denominator, 
    \begin{align*}
        \lt({\sum_{t=1}^T \E{h_t^2 \pi_t(a)^{-1}\Prob[a]{D\le T-t}} }\rt)^{\frac{2+\delta}{2}} \asymp T^{1+\frac{\delta}{2}}.
    \end{align*}
    Therefore, 
    \begin{align*}
        &\frac{\sum_{t\in[T]} h_t^{2+\delta}  \pi_t(a)^{-(1+\delta)}\Prob[a]{D\le T-t} }{\lt({\sum_{t=1}^T \E{h_t^2 \pi_t(a)^{-1}\Prob[a]{D\le T-t}} }\rt)^{\frac{2+\delta}{2}}}\\
        \le& \sum_{t\in[T]} t^{\frac{\alpha\delta}{2}}
        \asymp T^{\frac{-(1-\alpha)\delta}{2}} \to 0.
    \end{align*}
    \item For \ref{asp:variance-converge}, it is easy to check that with $h_t(a) = \sqrt{\pi_t(a)}$, we always have
    \begin{align*}
       \frac{{\sum_{t=1}^T {h_t^2 \pi_t(a)^{-1}\Prob[a]{D\le T-t}} }}{{\sum_{t=1}^T \E{h_t^2 \pi_t(a)^{-1}\Prob[a]{D\le T-t}} }} = 1.
    \end{align*}
\end{itemize}
\end{proof}

\section{Additional results}
\subsection{Why Hájek estimation is necessary for adjusting for delayed system?}
Regarding the question of why Hájek estimation is necessary for adjusting for delayed, even possibly never observed, outcomes, we hope to highlight that the key insight is that Hájek estimation accounts for arm-dependent delays. We first display both estimators (``H" for H\`{a}jek and ``NH" for non-H\`{a}jek):
    \begin{gather*}
    \hat{Q}_{H} = \frac{\sum_{t\in[T]} h_t(a)Y_t\mathbf{1}\{A_t = a, D_t\le T-t\}/\pi_t(a)}{\sum_{t\in[T]} h_t(a)\mathbf{1}\{A_t = a, D_t\le T-t\}/\pi_t(a)},\\
    \hat{Q}_{NH} = \frac{\sum_{t\in[T]} h_t(a)Y_t\mathbf{1}\{A_t = a, D_t\le T-t\}/\pi_t(a)}{\sum_{t\in[T]} h_t(a)\mathbf{1}\{D_t\le T-t\}}.
    \end{gather*}

    The numerators of both $\hat{Q}_{H}$ and $\hat{Q}_{NH}$ are the same, which has expectation 
    \begin{align*}
        \mu(a) \sum_{t\in[T]} h_t(a)\mathbb{P}_a(D_t\le T-t).
    \end{align*}
    Condition 2.3 and Lemma A.3 in the Appendix suggest the numerator has same order as
    \begin{align*}
        A = \mu(a) \cdot \mathbb{P}_a(D \le T) \cdot \sum_{t\in[T]} h_t(a).
    \end{align*}
    Similarly, we can prove that the denominator of $\hat{Q}_{H}$ has order
    \begin{align*}
        B_H = \mathbb{P}_a(D\le T) \cdot \sum_{t\in[T]} h_t(a),
    \end{align*}
    while the denominator of $\hat{Q}_{NH}$ has order
    \begin{align*}
        B_{NH} = \mathbb{P}(D\le T) \cdot \sum_{t\in[T]} h_t(a).
    \end{align*}
    We see that $A/B_H\sim \mu(a) $ but $A/B_{NH} \sim \mu(a)\mathbb{P}_a(D\le T)/\mathbb{P}(D\le T)$. Notably, the non-H\`{a}jek estimator does not distinguish arm-specific delay mechanisms. Therefore, H\`{a}jek estimation plays an important role in achieving better finite sample performance and is even necessary for asymptotic consistency when there are infinite delays (censoring) and $ \mathbb{P}_a(D < \infty)/\mathbb{P}(D < \infty) \neq 1$.

\subsection{Application of the results to other bandit algorithms}
Corollary \ref{cor:epsilon-greedy} suggests that as long as data collection policy $\pi_t(a)$ across all arms decays at a proper rate (slower than $O(t^{-1})$) and delay distribution has a thinner tail than $O(t^{-1/2})$, the assumptions \ref{asp:small-h}-\ref{asp:variance-converge} of our manuscript can be justified and the theorems of statistical inference can be applied.  To check the requirement for data collection policy in concrete bandit learning algorithms such as TS and UCB, the key step is to check how $\pi_t(a)$ evolves in $t$ in the implementation of the algorithm.

For TS, consider for simplicity a two-arm Beta-Bernoulli bandit, where arm $k$ is binary and has mean reward $\theta_k$. Apply independent Beta priors $\text{Beta}(\alpha_k, \beta_k)$ at the initial step. After $(T-1)$ steps, the posterior distribution is given by
        \begin{align*}
            P(\theta_0,\theta_1\mid \mathcal{H}_{t-1}) = \text{Beta}(\alpha_0 + N_T(0), \beta_0 + N_T(1))\text{Beta}(\alpha_1 + N_T(1), \beta_1 + N_T(0)),
        \end{align*}
where $N_T(0)$ and $N_T(1)$ are the number of observations on arm $0$ and $1$, respectively. At step $T$, the sampling probability for arm $1$ is given by 
        \begin{align*}
            \pi_T(1) = \mathbb{P}(\theta_1 \ge \theta_0 \mid \mathcal{H}_{t-1}).
        \end{align*}
In our inferential framework, we can take $h_t(1) = \sqrt{\pi_T(1)}$ and $h_t(0) = \sqrt{1- \pi_T(1)}$.
According to a probability result by Pham-Gia et al. (1993), the above formula has a closed form solution given by
        \begin{align*}
            \pi_T(1) = \frac{\text{B}(\alpha_1 + \alpha_0 + N_T-1, \beta_1 + \beta_0 + N_T-1)}{\text{B}(\alpha_1 + N_T(1), \beta_1 + N_T(0))\text{B}(\alpha_0 + N_T(1), \beta_0 + N_T(0))},
        \end{align*}
        where $\text{B}(\cdot,\cdot)$ is the Beta function so there is an explicit formula for $h_t(a)$.  

For inferential purposes, in practical implementation it might be helpful to add a clipping rate to the TS algorithm to avoid overly rapid decay in the sampling probabilities, as suggested by \citet{zhang2020inference}. Concretely, at step $t$, we sample arm $1$ with probability 
        \begin{align*}
            \pi_t(1) = \epsilon_t \vee \{(1-\epsilon_t) \wedge P(\theta_1\ge\theta_0\mid \mathcal{H}_{t-1})\}
        \end{align*} 
        where $\epsilon_t = Ct^{-\alpha}$ for some $\alpha\in[0,1)$. This will ensure arm 1 and arm 0 have sampling probability at least $Ct^{-\alpha}$, thus adequate for inference as implied by Corollary 4.1.

For UCB, it is a deterministic sampling policy and not directly applicable for inference purpose. We adopt the idea (similar to our side comment for TS) of adding a clipping rate that samples 
    \begin{align*}
        A_t = \left\{
        \begin{array}{cc}
           \arg\max_a \overline{Y}_{t-1}(a) + c\sqrt{\frac{\log_t}{N_t(a)}},  & \text{with probability } 1-\epsilon_t; \\
           \text{one of the rest arms,}  & \text{with probability } 1-\epsilon_t/(K-1);
        \end{array}
        \right.
    \end{align*}
Then the rate for $\pi_t(a)$ can also be controlled in a similar way.

\section{Additional numerical experiments}

\subsection{Simulation results under different delay mechanisms}

\begin{figure}[!ht]
    \centering
    \includegraphics[width = 0.5\columnwidth]{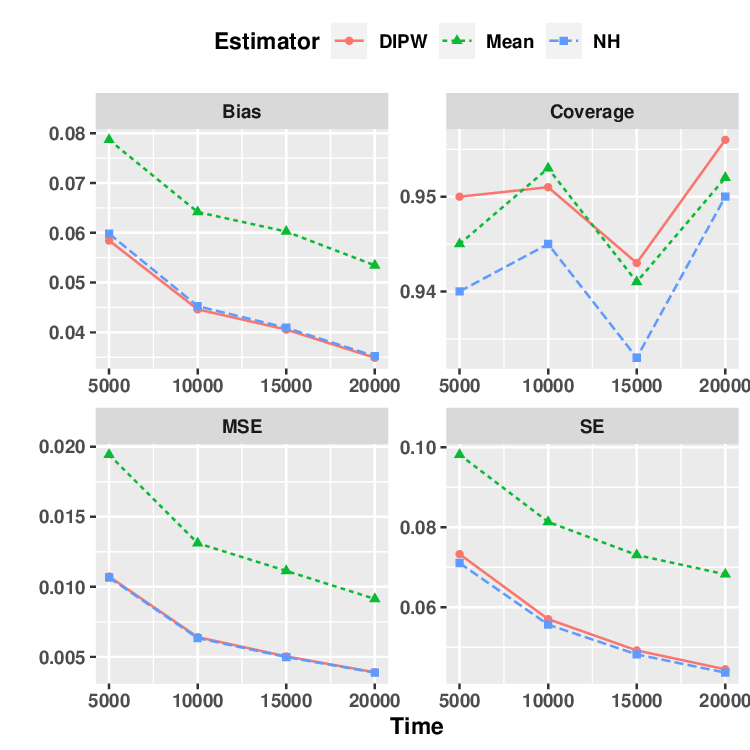}
    \caption{Evaluation of $\epsilon$-greedy with Negative Binomial delays}
    \label{fig:neg-delay}
\end{figure}

\begin{figure}[!ht]
    \centering
    \includegraphics[width = 0.5\columnwidth]{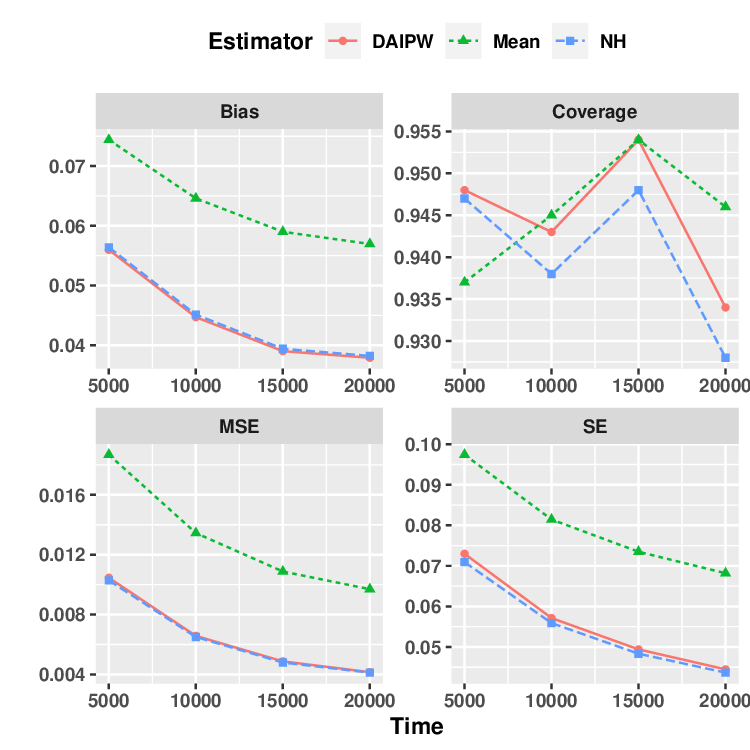}
    \caption{Evaluation of $\epsilon$-greedy with PARETO delays}
    \label{fig:pareto-delay}
\end{figure}

\subsection{Simulation results with different margins in $\epsilon$-greedy algorithms}

\begin{figure}[!ht]
    \centering
    \includegraphics[width = 0.5\columnwidth]{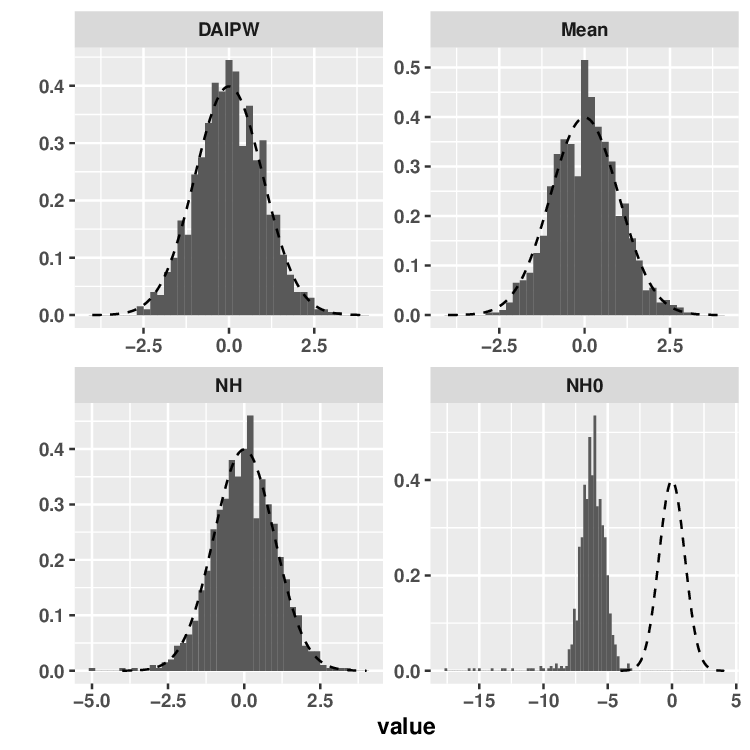}
    \caption{$\epsilon$-greedy over weak margin bandits $\mu(1) - \mu(2) = 0.1$. }
    \label{fig:weak-margin}
\end{figure}

\begin{figure}[!ht]
    \centering
    \includegraphics[width = 0.5\columnwidth]{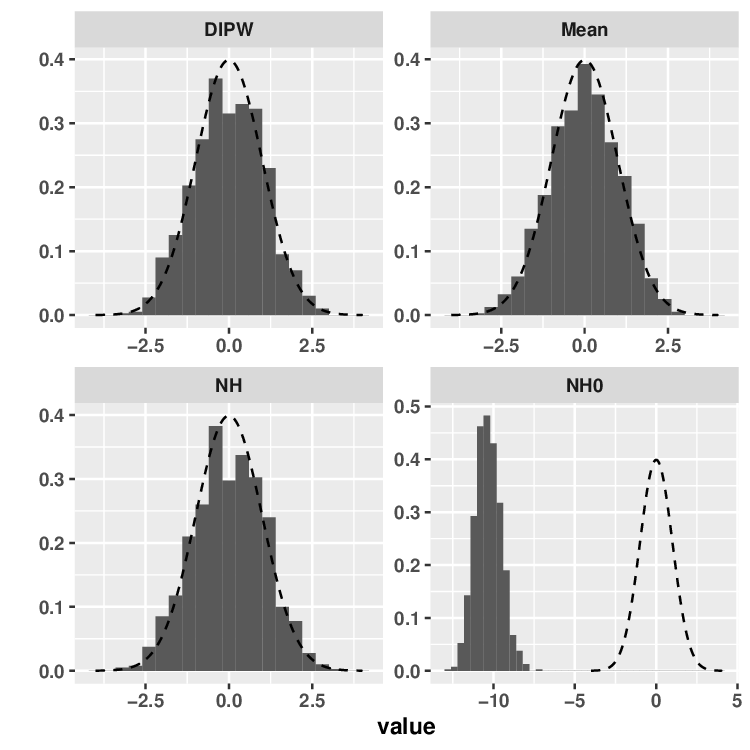}
    \caption{$\epsilon$-greedy over strong margin bandits $\mu(1) - \mu(2) = 0.5$. }
    \label{fig:strong-margin}
\end{figure}

\end{document}